\documentclass[12pt]{article}
\usepackage{amsmath}
\usepackage{txfonts,bm}  
\usepackage{cite}  
\usepackage{graphicx}
\newtheorem{theorem}{Theorem}[section]
\newtheorem{proposition}[theorem]{Proposition}
\newtheorem{lemma}[theorem]{Lemma}
\newtheorem{remark}[theorem]{Remark}
\newcommand{\qed}{\qquad$\square$}
\makeatletter
\@addtoreset{equation}{section}
\renewcommand{\theequation}{\@arabic\c@section.\@arabic\c@equation}
\long\def\@makecaption#1#2{
 \vskip 10pt 
 \setbox\@tempboxa\hbox{#1. #2}
 \ifdim \wd\@tempboxa >\hsize #1. #2\par \else \hbox
to\hsize{\hfil\box\@tempboxa\hfil} 
 \fi}
\makeatother
\begin{document}
\begin{center}
\renewcommand{\baselinestretch}{1.3}\selectfont
\begin{large}
\textbf{ULTRADISCRETIZATION OF A SOLVABLE TWO-DIMENSIONAL CHAOTIC MAP
 ASSOCIATED WITH THE HESSE CUBIC CURVE}
\end{large}\\[4mm]
\renewcommand{\baselinestretch}{1}\selectfont
\textrm{\large Kenji Kajiwara\footnote{kaji@math.kyushu-u.ac.jp},
Masanobu Kaneko\footnote{mkaneko@math.kyushu-u.ac.jp}, 
Atsushi Nobe$^\dag$\footnote{nobe@sigmath.es.osaka-u.ac.jp}
and Teruhisa Tsuda\footnote{tudateru@math.kyushu-u.ac.jp}}\\[2mm]
Faculty of Mathematics, Kyushu University,\\
6-10-1 Hakozaki, Fukuoka 812-8581, Japan\\[2mm]
$^\dag$ Graduate School of Engineering Science, Osaka University,\\
1-3 Machikaneyama-cho Toyonaka, Osaka 560-8531, Japan\\[4mm]
3 March 2009, Revised 6 March 2009
\end{center}
\begin{abstract}
We present a solvable two-dimensional piecewise linear chaotic map which
arises from the duplication map of a certain tropical cubic curve.  Its
general solution is constructed by means of the ultradiscrete theta
function.  We show that the map is derived by the ultradiscretization of
the duplication map associated with the Hesse cubic curve. We also show
that it is possible to obtain the nontrivial ultradiscrete limit of the
solution in spite of a problem known as ``the minus-sign problem.''
\end{abstract}
%


\section{Introduction}
Ultradiscretization\cite{TTMS:ultra} has been widely recognized as a powerful tool to extend the
theory of integrable systems to the piecewise linear discrete dynamical systems (ultradiscrete
systems)\cite{HT:uTzitzeica,IMNS:uSG,MSTTT:uToda,N:uToda,NT:uBurgers,TH:Permanent,TM:umKdV,TS:SCA,TTM:undKP,TH:uKdV,YYT:pBBS}.
In particular, it yields various soliton cellular automata when it is possible to discretize the
dependent variables into finite numbers of integers by a suitable choice of parameters.  It has also
established the links between the theory of integrable systems and various areas of mathematical
sciences, such as combinatorics, representation theory, tropical geometry, traffic flow models, and
so
on\cite{HHIK:AM_automata,HIK:crystal,Inoue-Takenawa,Iwao-Tokihiro:upToda,KT:upToda,KOSTY:KKR,KS:utheta,KSY:tau_combinatorialBethe,KTT:Bethe,MN:ELcorresponcende,NMT:u-2dBuergers,NT:uBurgers,Nobe:uQRT,TT:tropicalWeyl,TM:bb_and_Riemann,Tsuda:tropicalWeyl,Y:Amoebae}.
One of the key features of ultradiscretization is that one can obtain piecewise linear discrete
dynamical systems from rational discrete dynamical systems by a certain limiting procedure, which
corresponds to the low-temperature limit in the statistical mechanics. When this procedure is
applied to a certain class of discrete integrable systems, wide classes of exact solutions, such as
soliton solutions or periodic solutions, survive under the limit, which yield exact solutions to the
ultradiscrete integrable systems.

While the application of ultradiscretization to the integrable systems has achieved a great success,
it seems that only few results have been reported as to application to non-integrable
systems\cite{TM:OVmodel,TSU:pattern_formation}. One reason may be that in many cases of
non-integrable systems, fundamental properties are lost under the limit. For example, it is possible
to ultradiscretize the celebrated logistic map formally, but however, its chaotic behavior is lost through
the ultradiscretization\cite{KNT:1d}.

In \cite{KNT:1d}, the ultradiscretization of a one-dimensional chaotic map which arises as the
duplication formula of Jacobi's sn function has been considered. It exemplifies a {\it solvable
chaotic system}, which is regarded as a dynamical system lying on the border of integrability and
chaos\cite{GRV}, in the sense that though its exact solution is given by an elliptic function,
however, its dynamics exhibits typical chaotic behaviors such as irreversibility, sensitivity to
the initial values, positive entropy, and so on. By applying the ultradiscretization, it has been
shown that we obtain the tent map and its general solution simultaneously. Moreover, a tropical
geometric interpretation of the tent map has been presented, namely, it arises as the duplication
map on a certain tropical biquadratic curve. This result implies that there is the world of elliptic
curves and elliptic functions behind the tent map, which might be an unexpected and interesting
viewpoint. It also suggests that the tropical geometry and the ultradiscretization provides a
theoretical framework for the description of such a geometric aspect.

In this paper, we present two kinds of two-dimensional solvable chaotic maps and their general
solutions that are directly connected through the ultradiscretization. In Section 2, we construct a
piecewise linear map from a duplication map on a certain tropical plane cubic curve. We also construct
its general solution in terms of the ultradiscrete theta
function\cite{KNT:1d,KS:utheta,Mikhalkin-Zharkov,Nobe:utheta,Nobe:uQRT,TTGOR:uP} by using the
tropical Abel--Jacobi map. In Section 3, we consider a certain rational map which arises as a
duplication map on the Hesse cubic curve (see, for example,
\cite{AD:Hesse_pencil,JQ:Hessian,Smart:Hesse}), whose general solution is expressible in terms of
the theta functions of level 3.  In Section 4, we discuss the ultradiscretization of the rational
map and its solution obtained in Section 3, and show that they yield the piecewise linear map and
its solution obtained in Section 2. The rational map and its general solutions discussed in Section
3 and 4 involve a problem known as ``the minus-sign problem,'' which is usually regarded as an obstacle to
successful application of the ultradiscretization. We show that it is possible to overcome the
problem by taking careful parametrization and limiting procedure.
\section{Duplication map on tropical cubic curve}
\subsection{Duplication map}
In this section, we construct the duplication map on a certain tropical curve. For basic notions of the tropical geometry, we refer to
\cite{Gathmann,IMS:tropical_book,Mikhalkin:enumerative,Mikhalkin:application,Mikhalkin-Zharkov,RST:1st_step}.

\begin{figure}[h] 
\begin{center}
\begin{minipage}{0.3\textwidth}
\hfil \includegraphics[height=4cm]{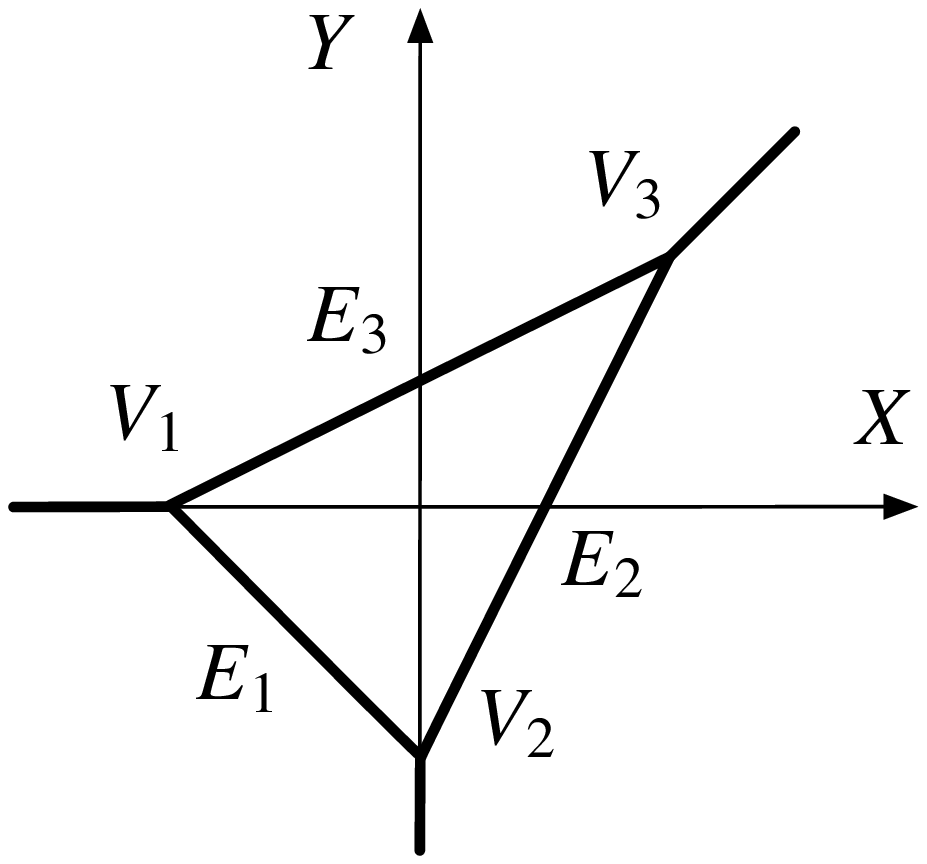}\hfil \\
\hfil (a) \hfil
\end{minipage}
\begin{minipage}{0.3\textwidth}
\hfil\includegraphics[height=4cm]{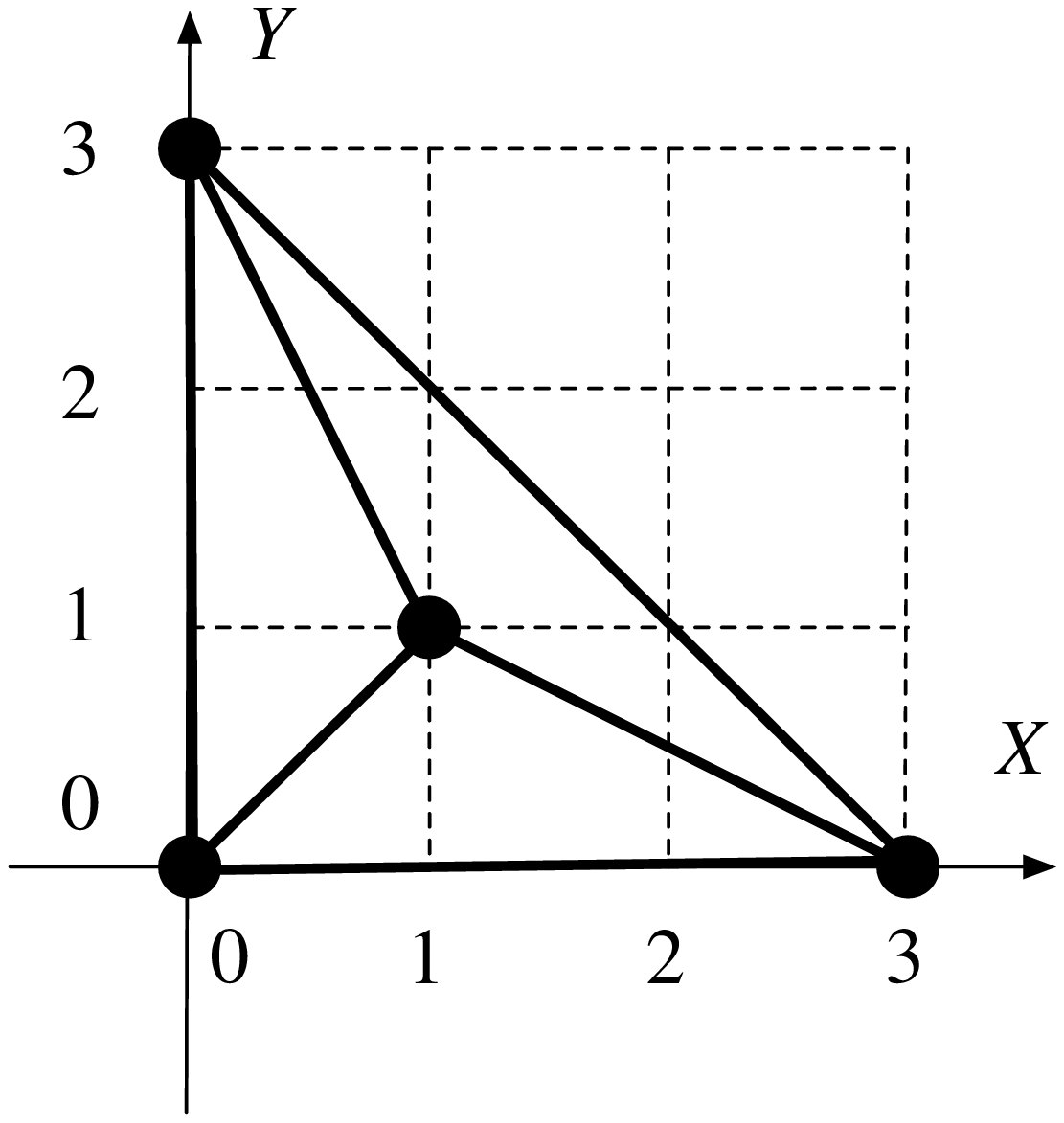} \hfil\\
\hfil (b) \hfil
\end{minipage}
\begin{minipage}{0.3\textwidth}
\hfil\includegraphics[height=4cm]{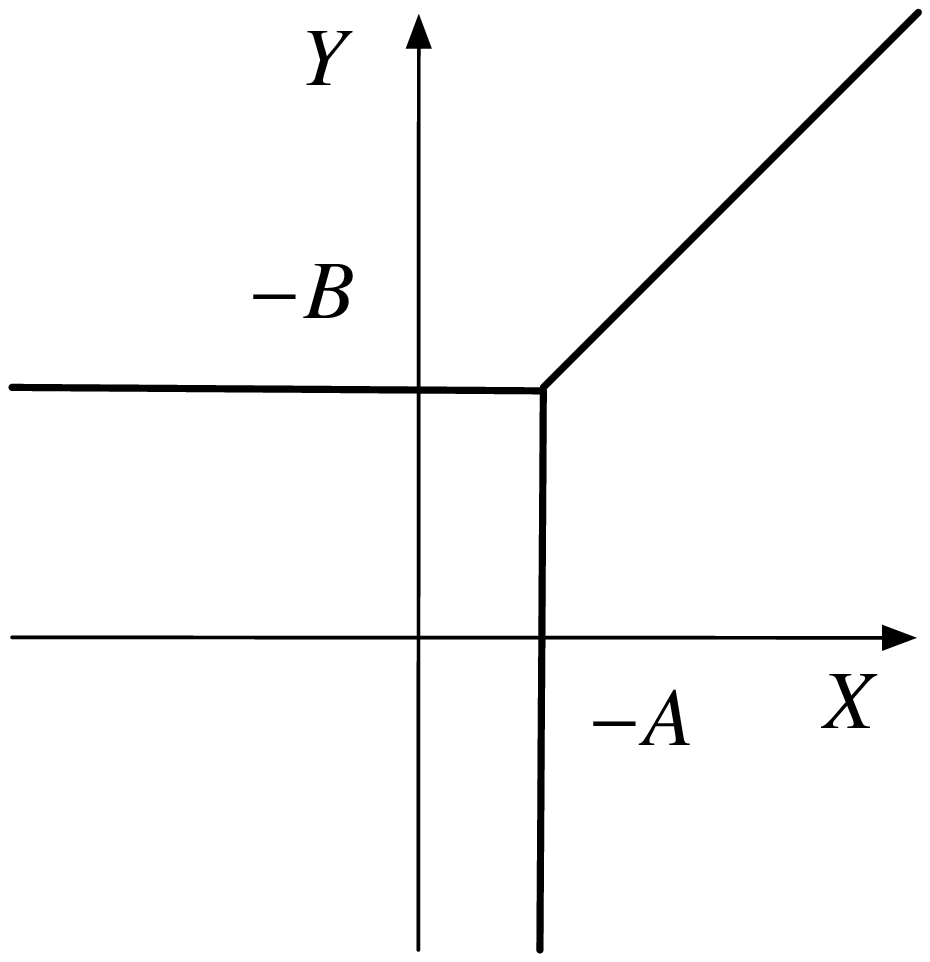} \hfil\\
\hfil (c) \hfil
\end{minipage}
\caption{(a): Tropical curve $C_K$. $V_1=(-K,0)$, $V_2=(0,-K)$, $V_3=(K,K)$. Primitive tangent vector
 of each edge: $E_1$: $\bm{v}_1=(1,-1)$, $E_2$: $\bm{v}_2=(1,2)$, $E_3$: $\bm{v}_3=(2,1)$. (b):
 Newton subdivision of the support of $C_K$. (c): Tropical line.}
 \label{fig:tropical_curve_support}
\end{center}
\end{figure}

Let us consider the tropical curve $C_K$ given by the tropical polynomial
\begin{equation}
 \Psi(X,Y;K)=\max\left[3X,3Y,X+Y+K,0\right],\quad X,Y,K\in\mathbb{R},\quad K>0.
\end{equation}
The curve $C_K$ is defined as the set of points where $\Psi$ is not differentiable. As shown in
Fig.\ref{fig:tropical_curve_support}(a), the vertices $V_i$ and the edges $E_i$ of $C_K$ are given by
$V_1=(-K,0)$, $V_2=(0,-K)$, $V_3=(K,K)$ and $E_1=V_1V_2$, $E_2=V_2V_3$, $E_3=V_3V_1$,
respectively. From the Newton subdivision of the support of $C_K$ given in
Fig.\ref{fig:tropical_curve_support}(b), we see that $C_K$ is a degree $3$ curve.  For a vertex on
the tropical curve, let $\bm{v}_i\in\mathbb{Z}^2$ ($i=1,\ldots,n$) be the primitive tangent vectors
along the edges emanating from the vertex. Then it is known that for any vertex there exist natural
numbers $w_i\in\mathbb{Z}_{>0}$ ($i=1,\ldots,n$) such that the following balancing condition holds:
\begin{equation}
 w_1\bm{v}_1+\cdots +w_n\bm{v}_n=(0,0).
\end{equation}
We call $w_i$ the weight of corresponding edge. Now, since the primitive tangent vectors emanating
from $V_1$ is given by $(-1,0)$, $(1,-1)$ and $(2,1)$, the balancing condition at $V_1$ is given by
\begin{equation}
 3(-1,0)+(1,-1)+(2,1)=(0,0).
\end{equation}
Therefore, the weight of $E_1$, $E_3$ and the tentacle along the edges emanating from $V_1$ are
given by $1$, $1$ and $3$, respectively. The balancing condition at $V_2$ and $V_3$ shows that the
weight of the edges $E_i$ ($i=1,2,3$) are $1$, and those of tentacles of $C_K$ are all $3$,
respectively. If a vertex $V$ is 3-valent, namely $V$ has exactly three adjacent edges whose
primitive tangent vectors and weights are $\bm{v}_i$ and $w_i$ ($i=1,2,3$), respectively, the
multiplicity of $V$ is defined by
$w_1w_2\left|\det(\bm{v}_1,\bm{v}_2)\right|=w_2w_3\left|\det(\bm{v}_2, \bm{v}_3)\right|
=w_3w_1\left|\det(\bm{v}_3, \bm{v}_1)\right|$.  If all the vertices of the tropical curve are
3-valent and have multiplicity $1$, then the curve is said to be smooth. The multiplicity of the
vertex $V_1$ is computed as
\begin{equation}
 3\cdot 1\cdot\left|~\det\left(\begin{array}{cc}-1 & 0\\1&-1 \end{array}\right)~\right|=3,
\end{equation}
and similarly those of $V_2$ and $V_3$ are both $3$, which imply that $C_K$ is not smooth.  The
genus is equal to the first Betti number of $C_K$, which is $1$ as shown in
Fig.\ref{fig:tropical_curve_support}(a).  Thus the curve $C_K$ is a non-smooth, degree $3$ tropical
curve of genus $1$.  Note that the cycle $\overline{C}_K$ of $C_K$ (the triangle obtained by
removing the tentacles from $C_K$) can be given by the equation
\begin{equation}
 \max[3X,3Y,0]=X+Y+K.\label{eqn:cycle}
\end{equation}

A tropical line is the tropical curve given by the tropical polynomial of the form
\begin{equation}
 L(X,Y)=\max[X+A,Y+B,0],
\end{equation}
which is shown in Fig.\ref{fig:tropical_curve_support} (c). The three primitive tangent vectors
emanating from the vertex are given by $(-1,0)$, $(0,-1)$, $(1,1)$. From the balancing condition
$(-1,0)+(0,-1)+(1,1)=(0,0)$, the weight of edges are all $1$.

Vigeland\cite{Vigeland} has introduced the group law on the tropical elliptic curve, which is a
smooth, degree 3 curve of genus 1. According to the group law, the duplication map is formulated as
follows; let $C$ be a tropical elliptic curve and let $\overline{C}$ be its cycle. Take a point
$P\in\overline{C}$. We draw a tropical line that intersects with $\overline{C}$ at $P$ with the
intersection multiplicity 2, and denote the other intersection point by $P*P$. Drawing a tropical
line passing through ${\cal O}$ and $P*P$ with a suitable choice of the origin of addition ${\cal
O}\in\overline{C}$, the third intersection point is $2P$.

For a given point $P$ on a tropical elliptic curve, the tropical line that intersects at $P$ with
the intersection multiplicity $2$ does not exist in general.  However, the curve $C_K$ has a
remarkable property that it is possible to draw a tropical line that intersects at any point on
$\overline{C}_K$ with the intersection multiplicity $2$. The explicit form of the duplication map is
given as follows:
\begin{proposition}\label{Prop:tropical_duplication}
Choosing the origin as ${\cal O}=V_3$, the duplication map $\overline{C}_K\ni P=(X,Y)\longmapsto
 2P=(\overline{X},\overline{Y})\in \overline{C}_K$ on the tropical cubic
 curve $C_K$ is given by
\begin{equation}
\overline{X}=Y+3\max[0,X]-3\max[X,Y],\quad \overline{Y}=X+3\max[0,Y]-3\max[X,Y],\label{map:u-Hesse}
\end{equation}
or 
\begin{equation}
X_{n+1}=Y_n+3\max[0,X_n]-3\max[X_n,Y_n],\quad Y_{n+1}=X_n+3\max[0,Y_n]-3\max[X_n,Y_n],\label{map:u-Hesse2}
\end{equation}
where $(X_n,Y_n)$ is the point obtained by the $n$ times successive applications of the map to $(X,Y)$.
\end{proposition}
\noindent\textbf{Proof.}\par
\begin{figure}[h] 
\begin{center}
\begin{minipage}{0.3\textwidth}
\hfil \includegraphics[height=4cm]{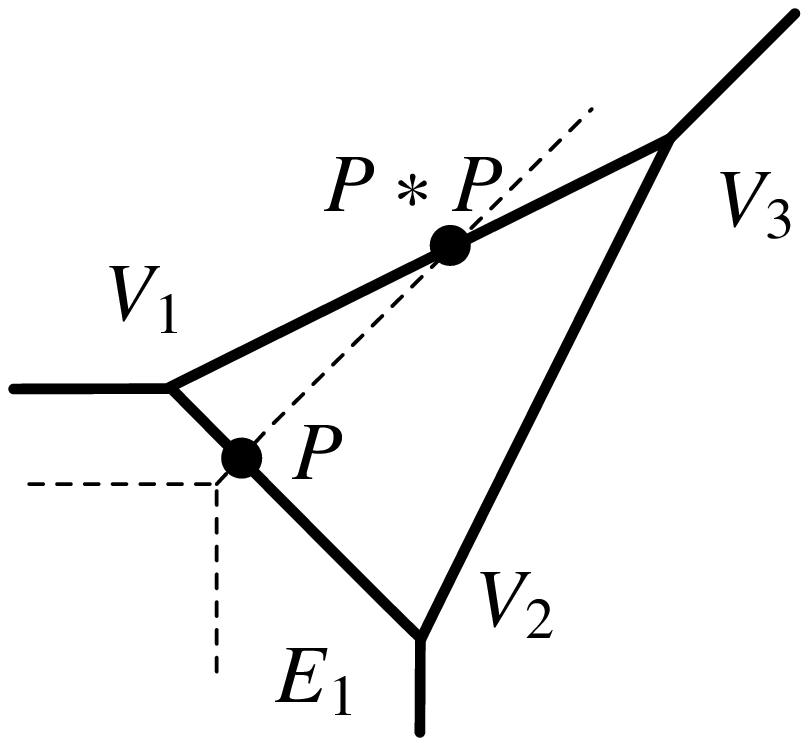}\hfil \\
\hfil (a) \hfil
\end{minipage}
\begin{minipage}{0.3\textwidth}
\hfil\includegraphics[height=4cm]{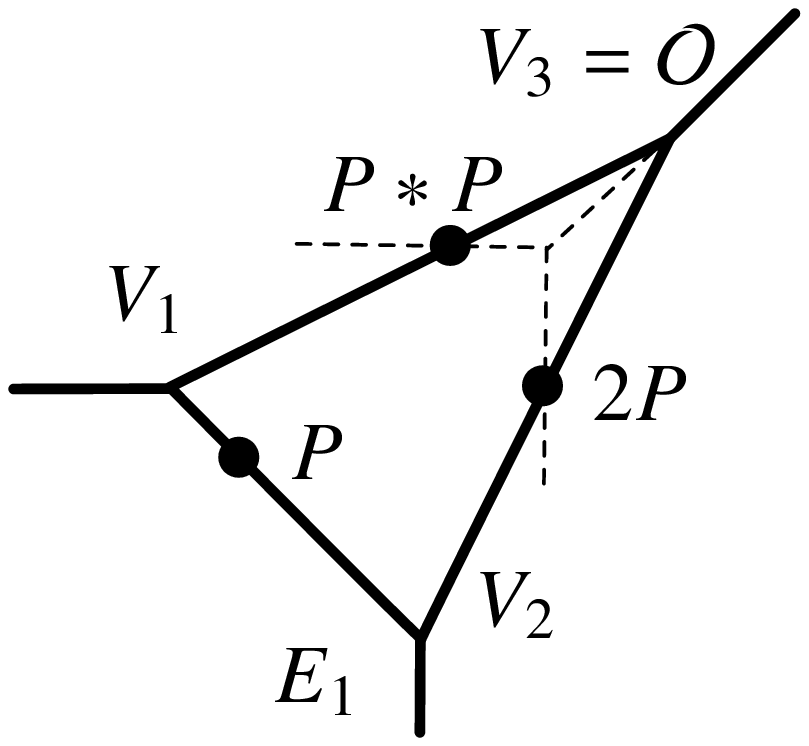} \hfil\\
\hfil (b) \hfil
\end{minipage}
\caption{(a): Map $P\mapsto P*P$ for $P\in E_1$. The intersection point of $C_K$ and the line
 passing through $P$ with multiplicity $2$ (broken line) is $P*P$. (b): Map $P*P\mapsto 2P$. The
 intersection point of $C_K$ and the line passing through ${\cal O}=V_3$ and $P*P$ (broken line) is
 $2P$. Obviously $P*P$ and $2P$ are symmetric with respect to $X=Y$.}  \label{fig:P_on_E1}
\end{center}
\end{figure}

\noindent\textbf{Case 1: $\bm{P\in E_1}$}\qquad As illustrated in Fig.\ref{fig:P_on_E1}(a), the
primitive tangent vectors of the two edges passing through $P$ are $(1,-1)$ (thick line) and $(1,1)$
(broken line), respectively, and the weight of the edges crossing at $P$ are both $1$. Then the
intersection multiplicity is given by
\begin{equation}
1\cdot 1\cdot \left|~\det\left(\begin{array}{cc}1 &-1 \\1 & 1 \end{array}\right)~\right|=2.
\end{equation}
Let $P*P$ be the other intersection point. Note that the intersection multiplicity at $P*P$ is
$1$. Then the map $P=(X,Y)\longmapsto P*P=(X',Y')$ is constructed as follows.  Since $P\in E_1$ and
$P*P\in E_1\cup E_2$, we have
\begin{equation}
 0=X+Y+K,\quad \max[3X',3Y']=X'+Y'+K.
\end{equation}
Subtracting the second equation from the first one, we obtain by using $(X-X')/(Y-Y')=1$
\begin{equation}
 X'=X-3\max[X,Y],\quad Y' = Y-3\max[X,Y].
\end{equation}
Our choice of the origin of addition ${\cal O}=V_3$ makes the form of $2P$ simple. It is obvious
as illustrated in Fig.\ref{fig:P_on_E1} (b), that $2P=(\overline{X},\overline{Y})$ is given by
$(\overline{X},\overline{Y})=(Y',X')$. Hence we obtain the map $P\longmapsto 2P$ as
\begin{equation}
 \overline{X}=Y-3\max[X,Y],\quad \overline{Y}=X-3\max[X,Y],\quad (X,Y)\in E_1.\label{map:P_on_E1}
\end{equation}
\begin{figure}[h] 
\begin{center}
\begin{minipage}{0.3\textwidth}
\hfil \includegraphics[height=4cm]{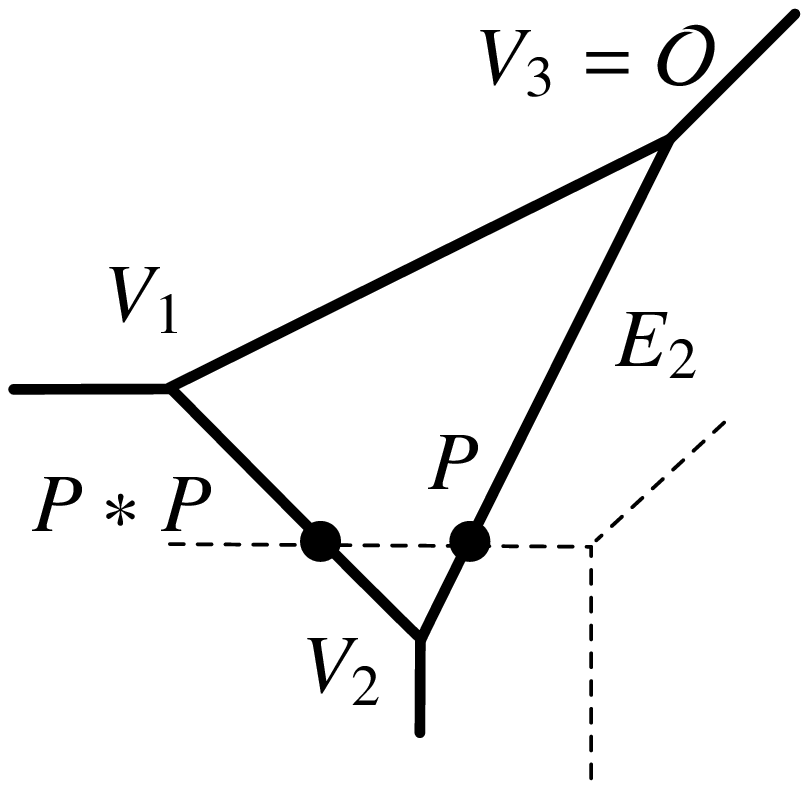}\hfil \\
\hfil (a) \hfil
\end{minipage}
\begin{minipage}{0.3\textwidth}
\hfil\includegraphics[height=4cm]{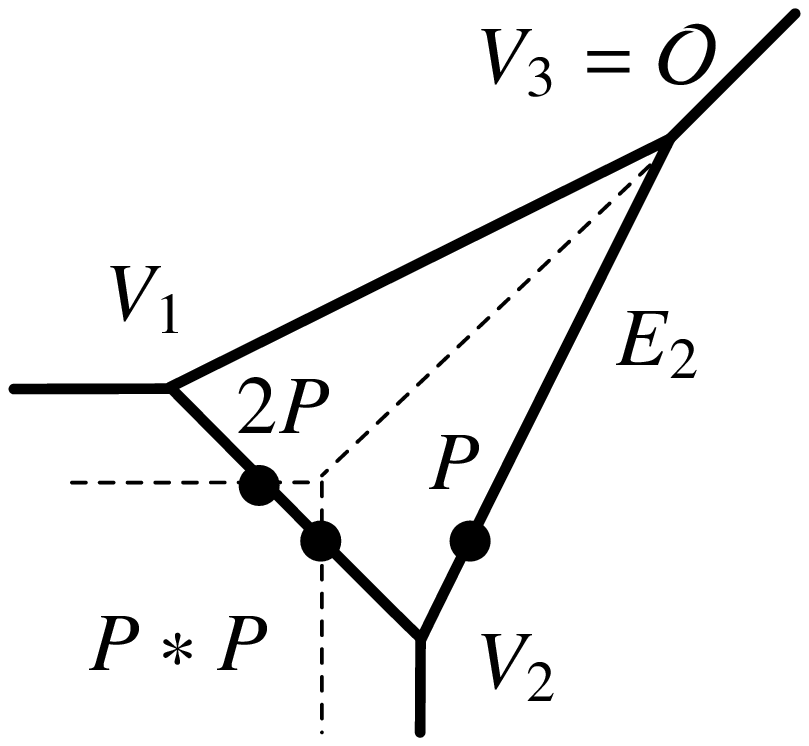} \hfil\\
\hfil (b) \hfil
\end{minipage}
\caption{(a): Map $P\mapsto P*P$ for $P\in E_2$. (b): Map $P*P\mapsto 2P$. }
 \label{fig:P_on_E2}
\end{center}
\end{figure}
\noindent\textbf{Case 2: $\bm{P\in E_2}$}\qquad As illustrated in Fig.\ref{fig:P_on_E2}(a),
the two primitive tangent vectors of the edges passing through $P$ are $(1,2)$ and $(1,0)$,
and the weight of the edges crossing at $P$ are both $1$. Then the intersection multiplicity is given by
\begin{equation}
1\cdot 1\cdot \left|~\det\left(\begin{array}{cc}1 &2 \\1 & 0 \end{array}\right)~\right|=2.
\end{equation}
Since $P\in E_2$ and $P*P\in E_1\cup E_3$, we have
\begin{equation}
 3X=X+Y+K,\quad \max[3Y',0]=X'+Y'+K.
\end{equation}
By using $Y'=Y$, we obtain the map $P\longmapsto P*P$ and $P\longmapsto 2P$ as
\begin{equation}
 X'=-2X+3\max[0,Y],\quad Y' = Y,
\end{equation}
\begin{equation}
 \overline{X}=Y,\quad \overline{Y}=-2X+3\max[0,Y],\quad (X,Y)\in E_2,\label{map:P_on_E2}
\end{equation}
respectively.
\begin{figure}[h] 
\begin{center}
\begin{minipage}{0.3\textwidth}
\hfil \includegraphics[height=4cm]{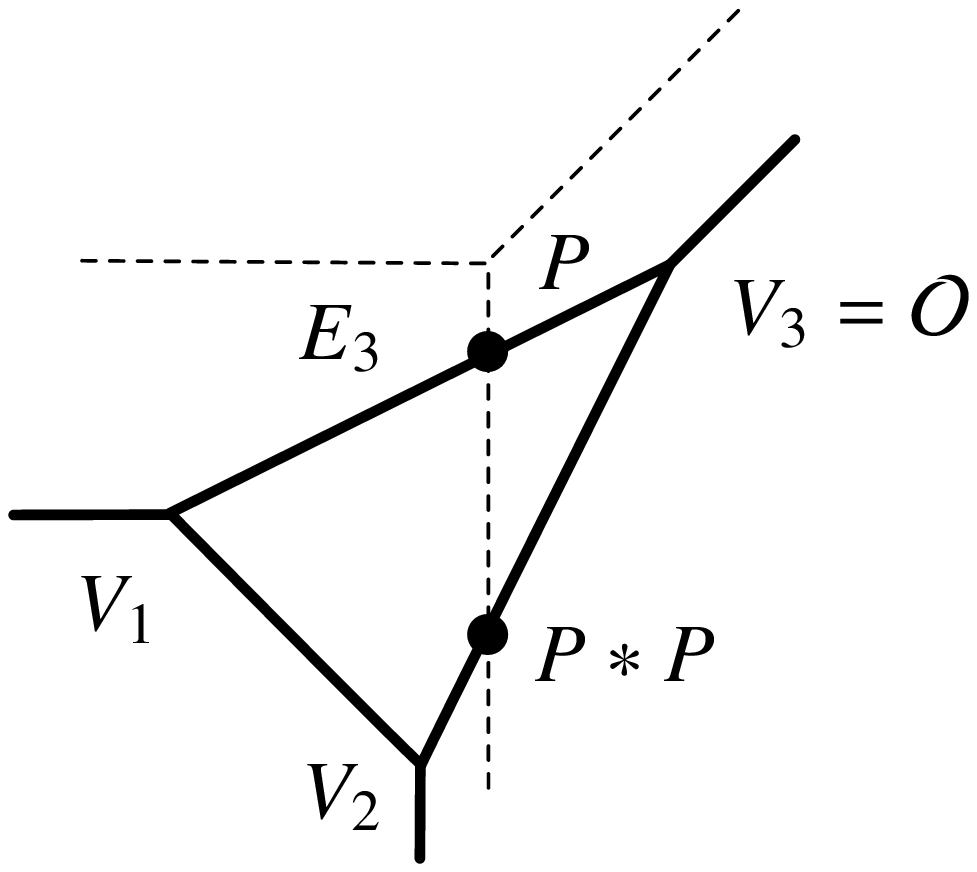}\hfil \\
\hfil (a) \hfil
\end{minipage}
\begin{minipage}{0.3\textwidth}
\hfil\includegraphics[height=4cm]{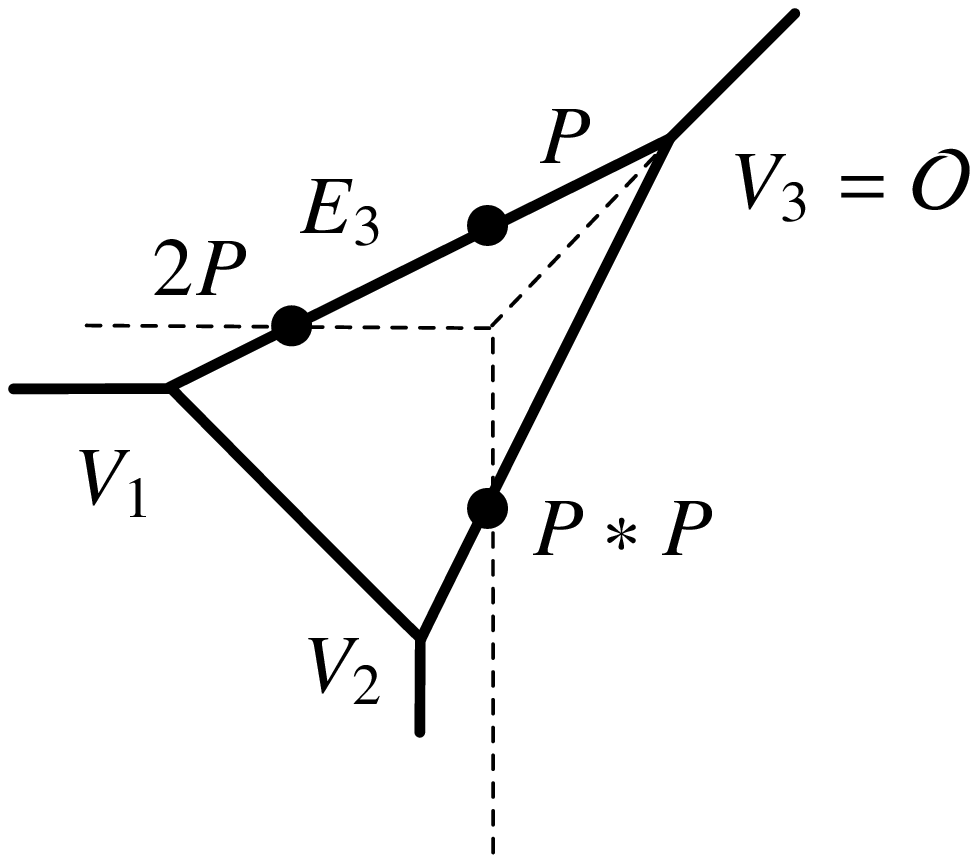} \hfil\\
\hfil (b) \hfil
\end{minipage}
\caption{(a): Map $P\mapsto P*P$ for $P\in E_3$. (b): Map $P*P\mapsto 2P$. }
 \label{fig:P_on_E3}
\end{center}
\end{figure}

\noindent\textbf{Case 3: $\bm{P\in E_3}$}\qquad As illustrated in Fig.\ref{fig:P_on_E3}(a),
the two primitive tangent vectors of the edges passing through $P$ are $(2,1)$ and $(0,1)$, 
and the weight of the edges crossing at $P$ are both $1$. The intersection multiplicity is given by
\begin{equation}
1\cdot 1\cdot \left|~\det\left(\begin{array}{cc}2 &1 \\0 & 1 \end{array}\right)~\right|=2.
\end{equation}
Since $P\in E_3$ and $P*P\in E_1\cup E_2$, we have
\begin{equation}
 3Y=X+Y+K,\quad \max[3X',0]=X'+Y'+K.
\end{equation}
By using $X'=X$, we obtain the map $P\mapsto P*P$ and $P\mapsto 2P$ as
\begin{equation}
 X'=X,\quad Y' = -2Y+3\max[0,X],
\end{equation}
\begin{equation}
\overline{X}=-2Y+3\max[0,X],\quad \overline{Y}=X, \quad (X,Y)\in E_3,\label{map:P_on_E3}
\end{equation}
respectively. 

We finally obtain eq.~(\ref{map:u-Hesse}) by collecting eqs.~(\ref{map:P_on_E1}), (\ref{map:P_on_E2}) and (\ref{map:P_on_E3}) together.\qed\\

Strictly speaking, the group law in \cite{Vigeland} cannot be applied to our case, since $C_K$ is
not a smooth curve. However, it is possible to show by direct computation that the map
(\ref{map:u-Hesse2}) is actually a duplication map on the tropical Jacobian of $\overline{C}_K$. For
this purpose, we first compute the total lattice length ${\cal L}$ of $\overline{C}_K$, which is
defined by the sum of the length of each edge scaled by the norm of corresponding primitive tangent
vector:
\begin{equation}
 {\cal L} = \sum_{i=1}^3 \frac{|E_i|}{|\bm{v}_i|} = \frac{\sqrt{5}K}{\sqrt{5}}+\frac{\sqrt{2}K}{\sqrt{2}}+\frac{\sqrt{5}K}{\sqrt{5}}=3K.
\end{equation}
Then the tropical Jacobian $J(\overline{C}_K)$ of $\overline{C}_K$ is given by
\begin{equation}
 J(\overline{C}_K)=\mathbb{R}/{\cal L}\mathbb{Z} = \mathbb{R}/3K\mathbb{Z}.
\end{equation}
The Abel--Jacobi map $\eta:~\overline{C}_K\rightarrow J(\overline{C}_K)$ is defined as the piecewise linear map satisfying
\begin{equation}
 \eta({\cal O})=\eta(V_3)=0,\quad \eta(V_1)=\frac{|E_3|}{|\bm{v}_3|}=K,\quad
\eta(V_2)=\eta(V_1)+\frac{|E_2|}{|\bm{v}_2|}=2K.
\end{equation}
\begin{proposition}\label{prop:tropical_Jacobian}
 The map $\overline{C}_K\ni P=(X,Y)\longmapsto
 \overline{P}=(\overline{X},\overline{Y})\in\overline{C}_K$ defined by eq.~(\ref{map:u-Hesse}) is a
 duplication map on the Jacobian $J(\overline{C}_K)$. Namely, we have $\eta(\overline{P})=2\eta(P)\ \mod 3K$.
\end{proposition}

\noindent\textbf{Proof.} \quad We consider the case $P\in E_1$. Suppose $P=(X,Y)$ satisfies
$V_1P:V_2P=s:1-s$ ($0\leq s\leq 1$), namely
\begin{equation}
 P=(X,Y)=(-(1-s)K,-sK),\quad \eta(P)=\eta(V_1)+sK=(1+s)K.
\end{equation}
\noindent Case (I): $X\leq Y$ ($0\leq s\leq\frac{1}{2}$). From eq.~(\ref{map:P_on_E1}), $\overline{P}$ is given by
\begin{equation}
 \overline{X}=Y-3Y=-2Y=2sK,\quad \overline{Y}=X-3Y=(-1+4s)K,\quad \overline{P}\in E_2,
\end{equation}
which implies
\begin{equation}
V_2\overline{P}:V_3\overline{P}=2s:1-2s,\quad \eta(\overline{P})=\eta(V_2)+2sK=2(1+s)K=2\eta(P).
\end{equation}
\noindent Case (II): $X\geq Y$ ($\frac{1}{2}\leq s\leq 1$). In this case, $\overline{P}$ is given by
\begin{equation}
 \overline{X}=Y-3X=(3-4s)K,\quad \overline{Y}=-2X=2(1-s)K,\quad \overline{P}\in E_3,
\end{equation}
which implies
\begin{equation}
V_3\overline{P}:V_1\overline{P}=-1+2s:2(1-s),\quad \eta(\overline{P})=(-1+2s)K\equiv
 2(1+s)K=2\eta(P) \mod 3K.
\end{equation}
Therefore we have shown that $\eta(\overline{P})=2\eta(P)$ for $P\in E_1$. We omit the proof of other
cases since they can be shown in a similar manner.\qed
\subsection{General solution}
\begin{figure}[h] 
\begin{center}
\begin{minipage}{0.4\textwidth}
\hfil \includegraphics[height=4cm]{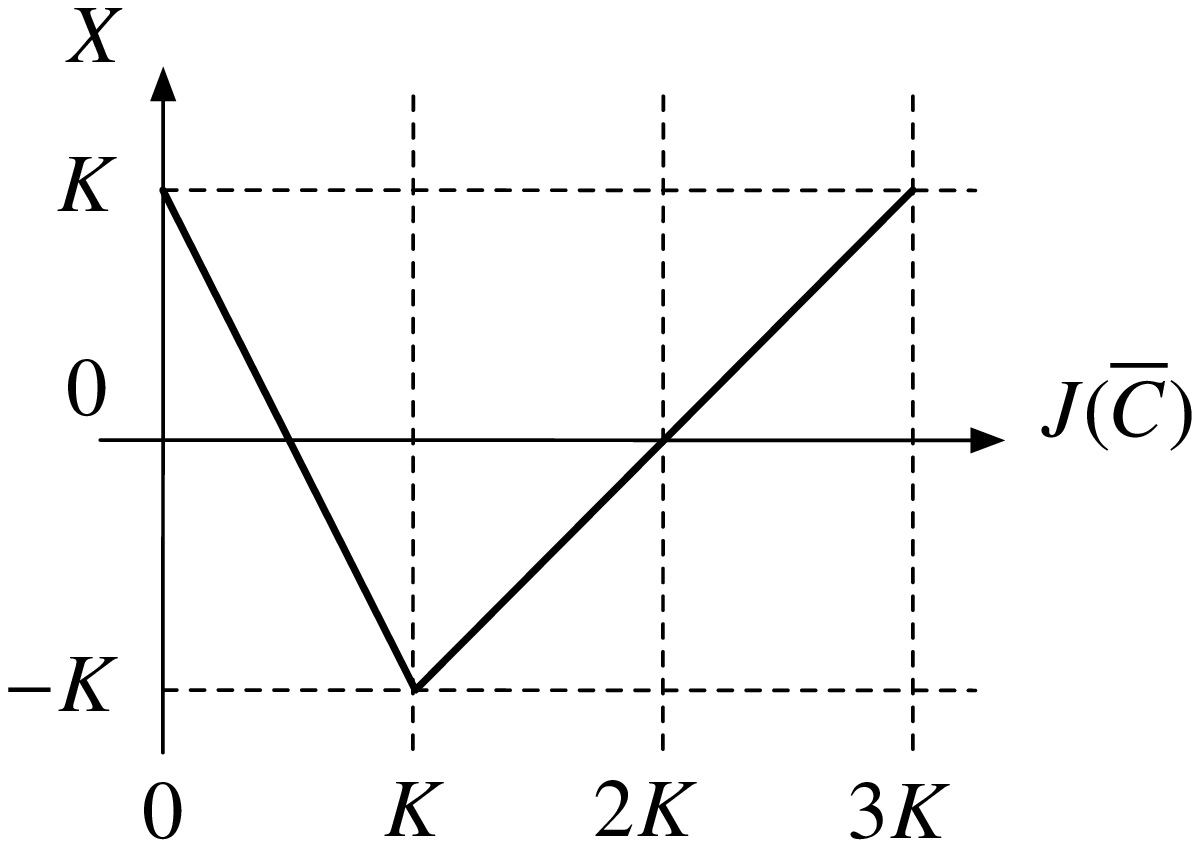}\hfil \\
\hfil (a) \hfil
\end{minipage}\qquad
\begin{minipage}{0.4\textwidth}
\hfil\includegraphics[height=4cm]{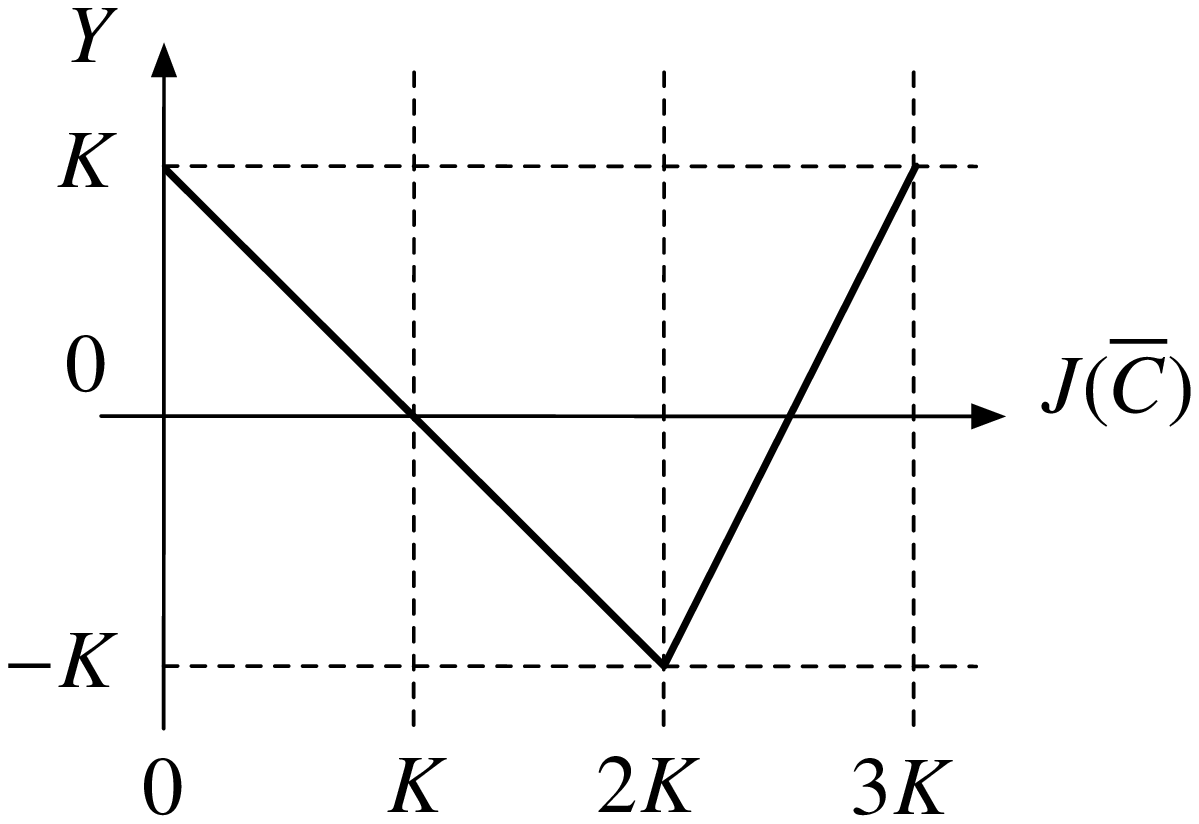} \hfil\\
\hfil (b) \hfil
\end{minipage}
\caption{(a): $\pi_1\circ\eta^{-1}:~J(\overline{C}_K)\rightarrow X$. (b): $\pi_2\circ\eta^{-1}:~J(\overline{C}_K)\rightarrow Y$.}
 \label{fig:projections}
\end{center}
\end{figure}

From the construction of the map (\ref{map:u-Hesse2}), it is possible to obtain the general solution
by using the Abel-Jacobi map of $\overline{C}_K$. Let $\pi_1$ and $\pi_2$ be projections from
$\overline{C}_K$ to the $X$-axis and the $Y$-axis, respectively. Then the maps $\pi_1\circ
\eta^{-1}$ and $\pi_2\circ \eta^{-1}$, namely, the maps from the tropical Jacobian to the $X$-axis and
the $Y$-axis through the Abel-Jacobi map are given as illustrated in Fig. \ref{fig:projections}(a) and (b),
respectively.
Therefore, Proposition \ref{prop:tropical_Jacobian} implies that
$X_n=\pi_1\circ \eta^{-1}(2^n u_0)$, $Y_n=\pi_2\circ \eta^{-1}(2^n u_0)$ for
arbitrary $u_0\in J(\overline{C}_K)$ gives the general solution to eq.~(\ref{map:u-Hesse2}).

\begin{figure}
 \begin{center}
  \includegraphics[height=4cm]{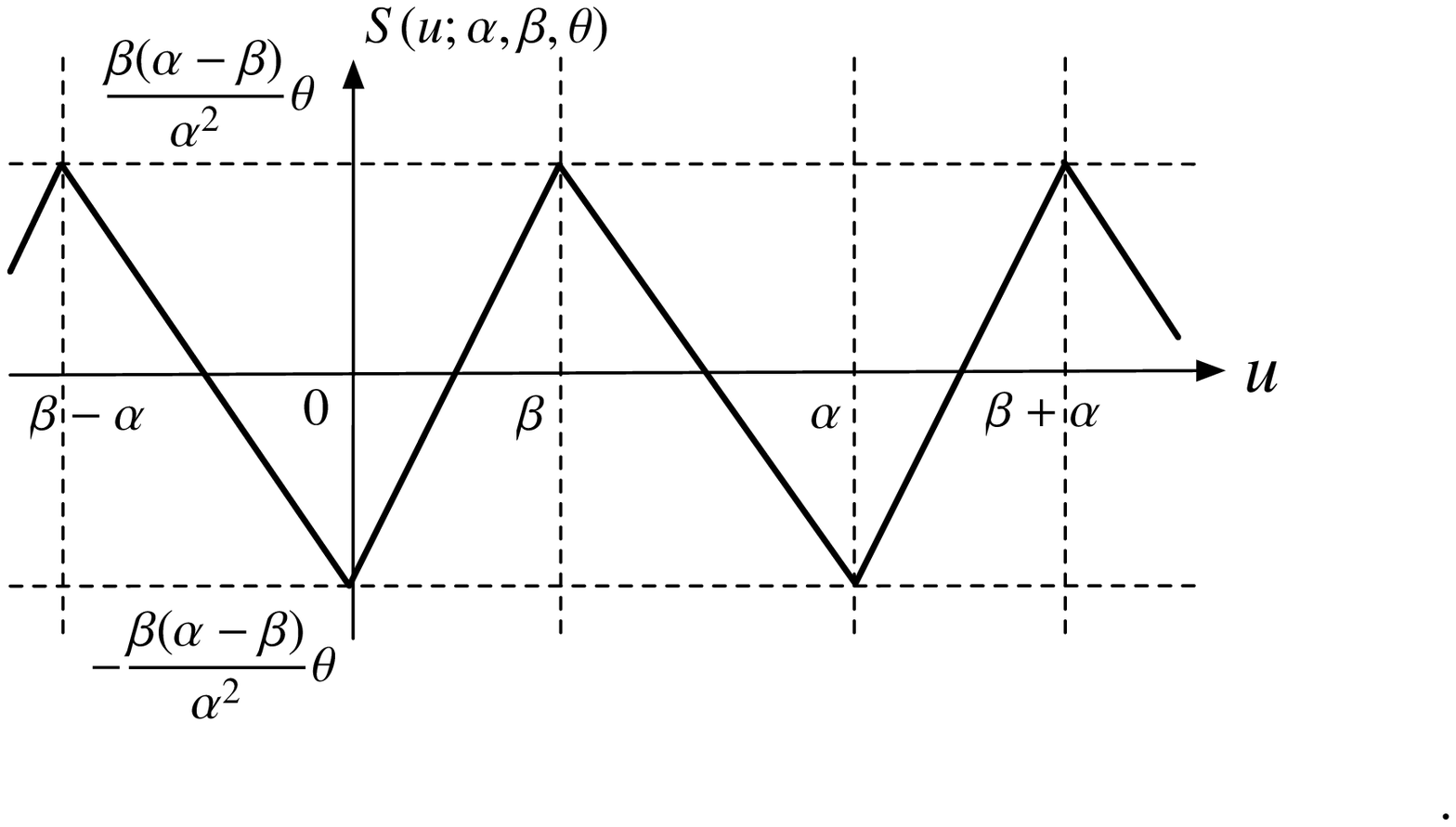}
 \end{center}
\caption{Graph of $S(u;\alpha,\beta,\theta)$.}\label{fig:S}
\end{figure}

It is possible to express $\pi_1\circ \eta^{-1}$ and $\pi_2\circ \eta^{-1}$ by using the
ultradiscrete theta function $\Theta(u;\theta)$ defined by\cite{KNT:1d,KS:utheta,Mikhalkin-Zharkov,Nobe:utheta,Nobe:uQRT,TTGOR:uP}
\begin{equation}
 \Theta(u;\theta)=-\theta\left\{((u))-\frac{1}{2}\right\}^2,\quad  ((u))=u-{\rm Floor}~(u).\label{eqn:u-theta}
\end{equation}
For this purpose, we introduce a piecewise linear periodic function $S(u;\alpha,\beta,\theta)$ by
\begin{equation}
 S(u;\alpha,\beta,\theta)=\Theta\left(\frac{u}{\alpha};\theta\right)-\Theta\left(\frac{u-\beta}{\alpha};\theta\right),\label{eqn:S}
\end{equation}
which has a period $\alpha$ and amplitude $2\beta(\alpha-\beta)\theta/\alpha^2$ as illustrated in Fig.\ref{fig:S}.
Comparing Fig. \ref{fig:projections} with Fig. \ref{fig:S}, we have
\begin{equation}
 \pi_1\circ \eta^{-1}(u) = S\left(u-K;3K,2K,\frac{9}{2}K\right),\quad
 \pi_2\circ \eta^{-1}(u) = S\left(u-2K;3K,K,\frac{9}{2}K\right).
\end{equation}
Therefore, we obtain the following proposition:
\begin{proposition}\label{prop:tropical_general_sol}
For a given initial value $P_0=(X_0,Y_0)$, the general solution to the map
(\ref{map:u-Hesse2}) is given by
\begin{equation}
\begin{array}{l}\medskip
 {\displaystyle  X_n = S\left(2^n u_0-K;3K,2K,\frac{9}{2}K\right),\quad
 Y_n = S\left(2^n u_0-2K;3K,K,\frac{9}{2}K\right),}\\
{\displaystyle K=3\max[X_0,Y_0,0]-X_0-Y_0,\quad u_0=\eta(P_0).}
\end{array}\label{gen_sol:u-Hesse}
\end{equation} 
\end{proposition}

\begin{figure}[h] 
\begin{center}
\begin{minipage}{0.4\textwidth}
\hfil \includegraphics[height=4cm]{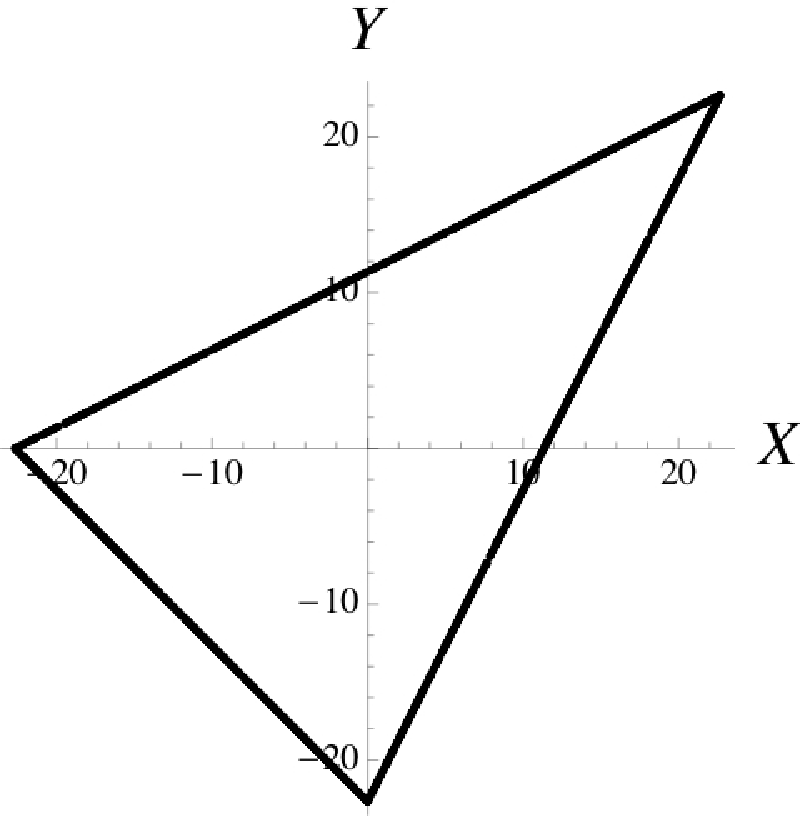}\hfil 
\end{minipage}
\begin{minipage}{0.4\textwidth}
\hfil\includegraphics[height=2.5cm]{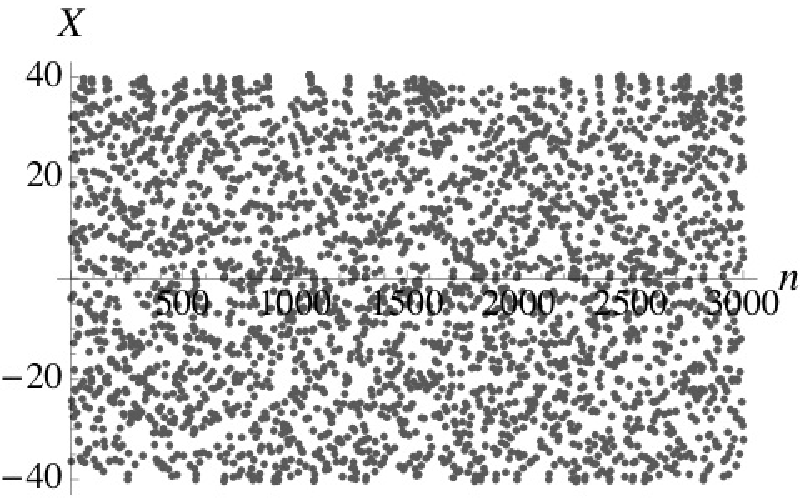} \hfil\\
\hfil\includegraphics[height=2.5cm]{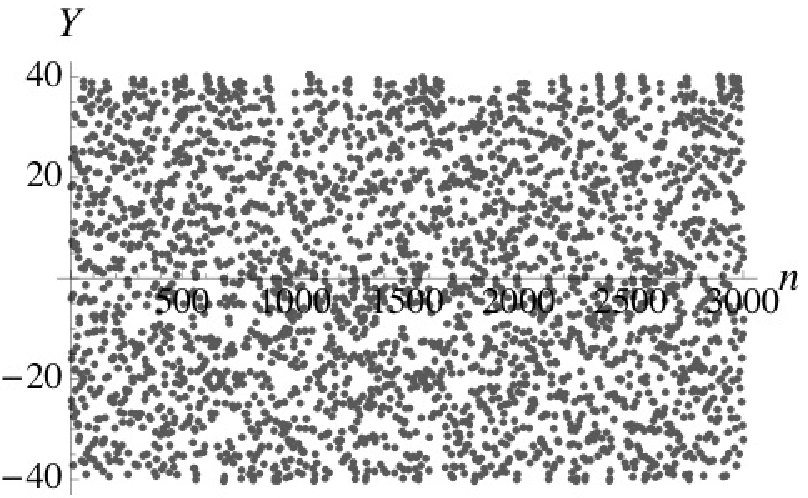} \hfil
\end{minipage}
\caption{Orbit of the map (\ref{map:u-Hesse2}) with the initial value $(X_0,Y_0)=(8.56546, 15.6231)$.}
 \label{fig:XY}
\end{center}
\end{figure}

Fig. \ref{fig:XY} shows the orbit of the map (\ref{map:u-Hesse2})
plotted with 3,000 times iterations.
The map has an invariant curve $\overline{C}_K$ given by eq.~(\ref{eqn:cycle}),
and the figure shows that the curve is filled with the points of the orbit.
\section{Duplication map on Hesse cubic curve}
\subsection{Duplication map}
The Hesse cubic curve is a curve in $\mathbb{P}^2$ given by
\begin{equation}
E_\mu :\quad x^3 + y^3 + 1 = 3\mu xy,\label{eqn:Hesse_curve}
\end{equation}
or in the homogeneous coordinates $[x_0:x_1:x_2]=[x:y:1]$
\begin{equation}
E_\mu:\quad x_0^3 + x_1^3 + x_2^3 = 3\mu x_0x_1x_2.\label{eqn:Hesse_curve_hom}
\end{equation}
The nine inflection points are given by $[1:-1:0]$, $[1:-\omega:0]$, $[1:-\omega^2:0]$, $[1:0:-1]$,
$[1:0:-\omega]$, $[1:0:-\omega^2]$, $[0:1:-1]$, $[0:1:-\omega]$, $[0:1:-\omega^2]$, where $\omega$
is a nontrivial third root of $1$.  It is known that any non-singular plane cubic curve is
projectively equivalent to $E_\mu$ (see, e.g.\cite{AD:Hesse_pencil}). Moreover, these inflection
points are also the base points of the pencil
\begin{equation}
 t_0(x_0^3 + x_1^3 + x_2^3) = t_1 x_0x_1x_2,\quad [t_0:t_1]\in\mathbb{P}^1.\label{eqn:Hesse_pencil}
\end{equation}

\begin{figure}[h] 
\begin{center}
\begin{minipage}{0.4\textwidth}
\hfil \includegraphics[height=4cm]{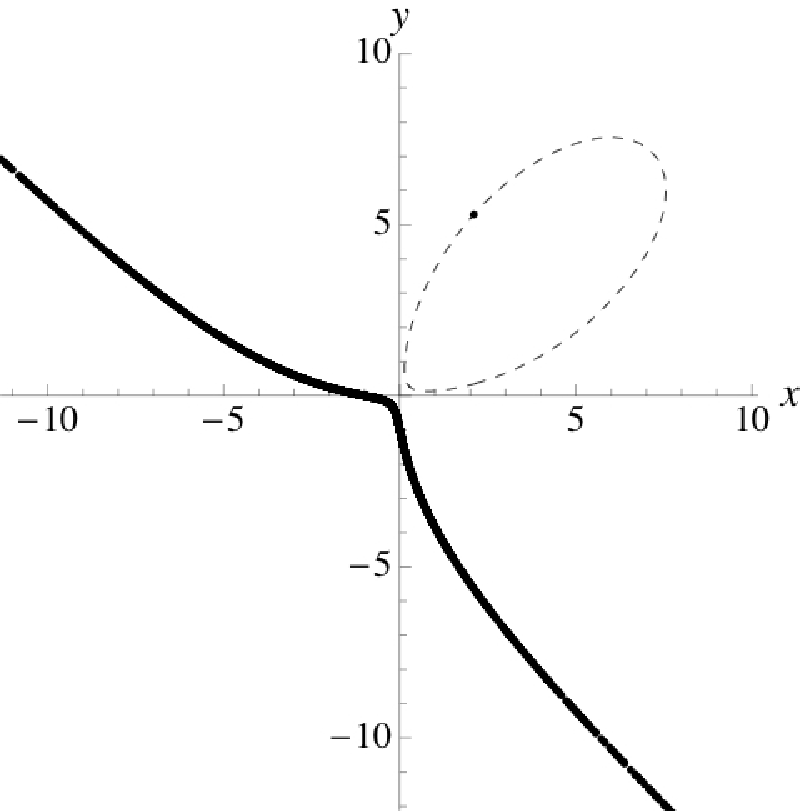}\hfil 
\end{minipage}
\begin{minipage}{0.4\textwidth}
\hfil\includegraphics[height=2.5cm]{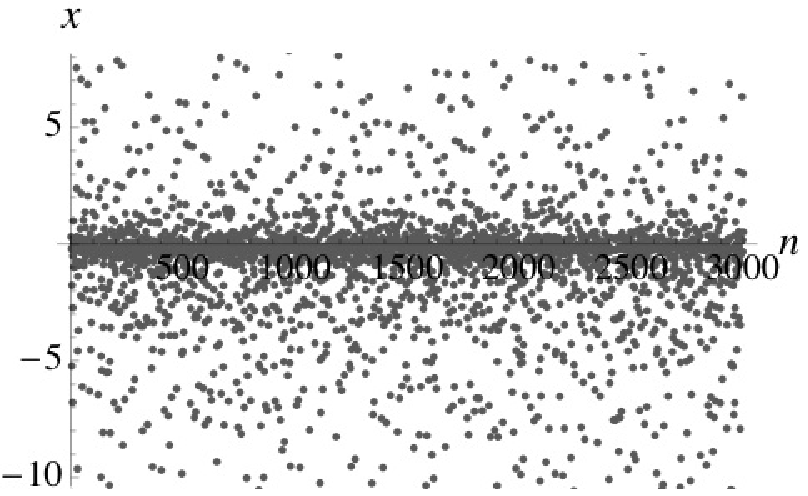} \hfil\\
\hfil\includegraphics[height=2.5cm]{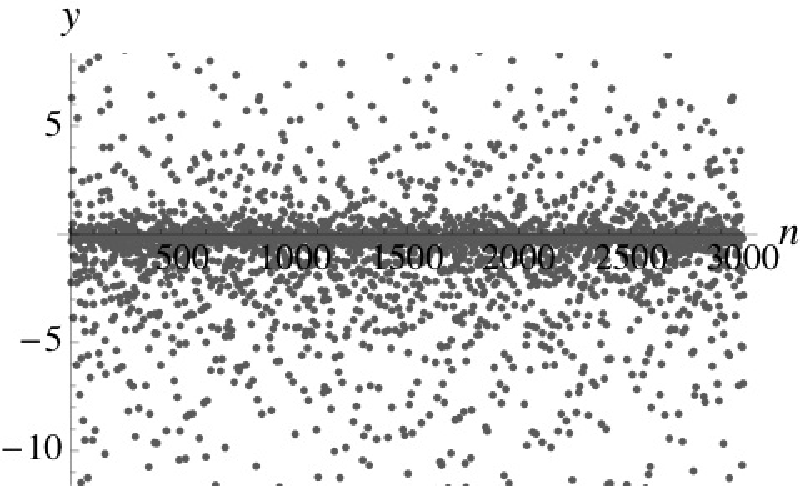} \hfil
\end{minipage}
\caption{Orbit of the map (\ref{map:Hesse2}) with the initial value $(x_0,y_0)=(2.1, 5.3)$. Dashed
 line of the left figure is the invariant curve.}
 \label{fig:Hesse_XY}
\end{center}
\end{figure}

\begin{figure}[h] 
\begin{center}
\begin{minipage}{0.4\textwidth}
\hfil \includegraphics[height=4cm]{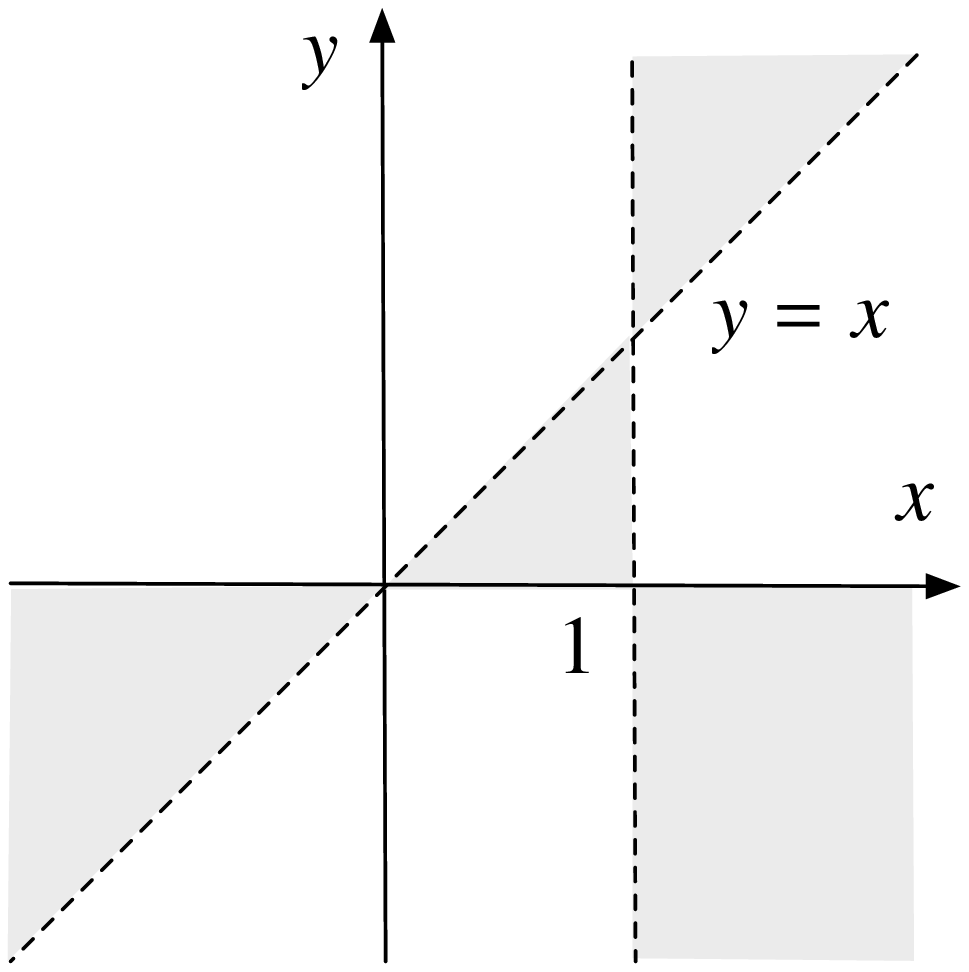}\hfil \\
\hfil (a) \hfil
\end{minipage}\qquad
\begin{minipage}{0.4\textwidth}
\hfil\includegraphics[height=4cm]{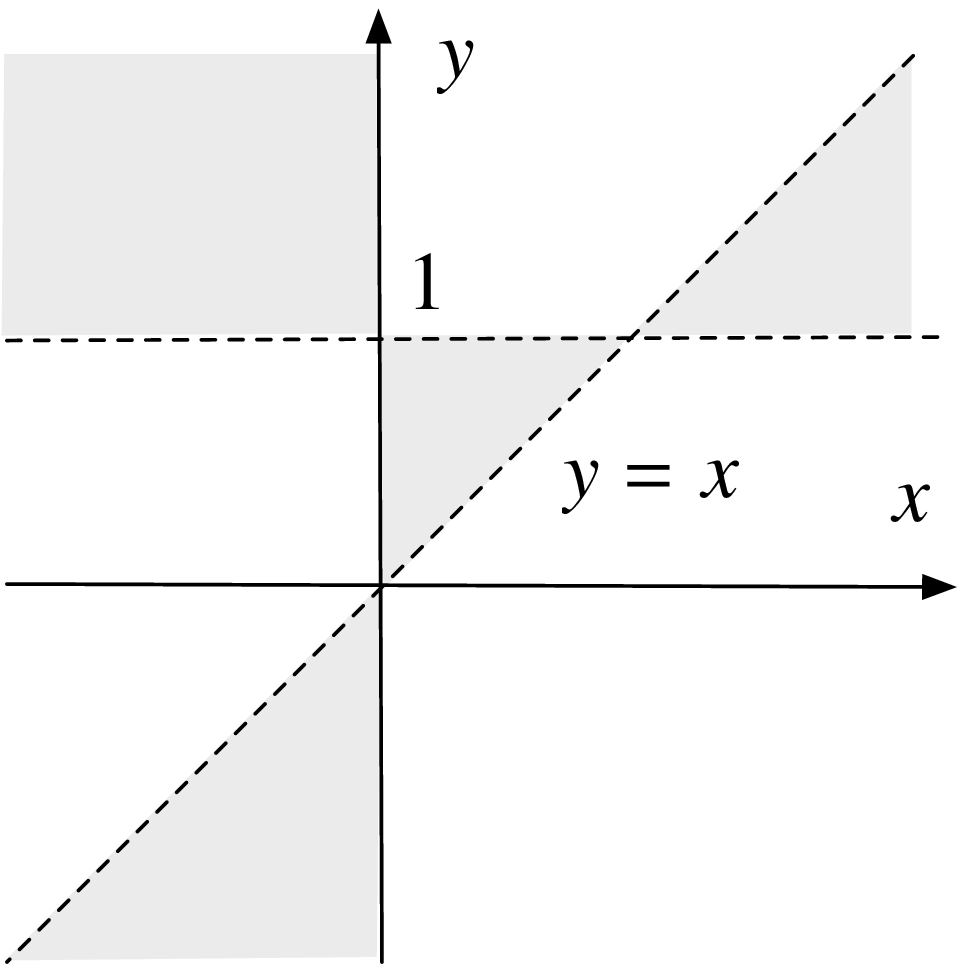} \hfil\\
\hfil (b) \hfil
\end{minipage}
\caption{(a): Region of $(x_{n-1},y_{n-1})$ for $x_n>0$. (b): Region of $(x_{n-1},y_{n-1})$  for $y_n>0$.}
 \label{fig:xy>0}
\end{center}
\end{figure}

The duplication map is constructed by the standard procedure; for an arbitrary point on $P\in E_\mu$
draw a tangent line, and set the other intersection of the tangent line and $E_\mu$ as $P*P$.  Taking
one of the inflection points as an origin ${\cal O}$ of addition, the intersection of $E_\mu$ and
the line connecting $P*P$ and ${\cal O}$ gives $2P$. Choosing ${\cal O}$ to be $[1:-1:0]$ among nine
inflection points of $E_\mu$, the duplication map $P=(x,y)\mapsto 2P=(\overline{x},\overline{y})$ is
explicitly calculated as (see, for example, \cite{JQ:Hessian,Smart:Hesse})
\begin{equation}
 \overline{x} = \frac{(1-x^3)y}{x^3-y^3},\quad \overline{y}=\frac{(1-y^3)x}{y^3-x^3},\label{map:Hesse}
\end{equation}
or writing the point obtained by the $n$ times applications of the map to $(x,y)$ as $(x_n,y_n)$, we have
\begin{equation}
  x_{n+1} = \frac{(1-x_n^3)y_n}{x_n^3-y_n^3},\quad y_{n+1}=\frac{(1-y_n^3)x_n}{y_n^3-x_n^3}.\label{map:Hesse2}
\end{equation}
By construction, it is obvious that the map (\ref{map:Hesse2}) has the invariant curve $E_\mu$,
where $\mu$ is the conserved quantity.  Fig. \ref{fig:Hesse_XY} shows the orbit of the map
(\ref{map:Hesse2}) plotted with 3,000 times iterations. Note that, although the invariant curve has
a component in the first quadrant $x,y>0$ for $\mu>0$, the real orbit never enters in this
quadrant (except for the initial point), which can be verified by a simple consideration; suppose
that $x_n>0$ at some $n$.  Then eq.~(\ref{map:Hesse2}) implies that $(x_{n-1},y_{n-1})$ must be in
the highlighted region of Fig. \ref{fig:xy>0}(a). On the other hand, if $y_n>0$ at some $n$,
$(x_{n-1},y_{n-1})$ must be in the highlighted region of Figure \ref{fig:xy>0}(b). Since the
intersection of the two regions is empty, it is impossible to realize $x_n,y_n>0$ for any $n$ as
long as we start from the real initial value.
\subsection{General solution}
The general solution to the map (\ref{map:Hesse}) or (\ref{map:Hesse2}) is given in terms of the
following theta functions of level $3$. Let us introduce the functions $\theta_k(z,\tau)$
$(k=0,1,2)$ by
\begin{equation}
 \theta_k(z,\tau)
=\sum_{n\in\mathbb{Z}} e^{3\pi i (n+\frac{k}{3}-\frac{1}{6})^2\tau} e^{6\pi i
(n+\frac{k}{3}-\frac{1}{6})(z+\frac{1}{2})}
=\vartheta_{\left(\frac{k}{3}-\frac{1}{6},\frac{3}{2}\right)}(3z,3\tau),\label{fn:theta_k}
\end{equation}
where $\vartheta_{\left(a,b\right)}(z,\tau)$ is the theta function with characteristic $(a,b)$
defined by
\begin{equation}
 \vartheta_{(a,b)}(z,\tau) 
= \sum_{n\in\mathbb{Z}} e^{\pi i (n+a)^2\tau+2\pi i(n+a)(z+b)},\quad 
\tau\in {\cal H}=\{{\rm Im}~z>0,\ z\in\mathbb{C}\}.\label{fn:theta_characteristic}
\end{equation}
\begin{proposition}\label{prop:general_sol_Hesse}
 The general solution to eq.~(\ref{map:Hesse2}) is given by
\begin{equation}
x_n = \frac{\theta_0(2^nz_0,\tau)}{\theta_2(2^nz_0,\tau)},\quad 
y_n = \frac{\theta_1(2^nz_0,\tau)}{\theta_2(2^nz_0,\tau)},\label{eqn:gen_sol:Hesse}
\end{equation}
where $z_0\in\mathbb{C}$ is an arbitrary constant.
\end{proposition}
Proposition \ref{prop:general_sol_Hesse} is a direct consequence of the following proposition:
\begin{proposition}\label{prop:theta}\hfill
 \begin{enumerate}
  \item $\theta_k(z,\tau)$ $(k=0,1,2)$ satisfy
\begin{equation}
 \theta_0(z,\tau)^3 +  \theta_1(z,\tau)^3 +  \theta_2(z,\tau)^3 = 3\mu(\tau)~\theta_0(z,\tau)\theta_1(z,\tau)\theta_2(z,\tau),\quad
\mu(\tau) = -\frac{\varphi'(0,\tau)}{\psi'(0,\tau)},\label{eqn:th_Hesse}
\end{equation}
where
\begin{equation}
\varphi(z,\tau)=\frac{\theta_1(z,\tau)}{\theta_0(z,\tau)},\quad
\psi(z,\tau)=\frac{\theta_2(z,\tau)}{\theta_0(z,\tau)}. \label{eqn:elliptic_fn}
\end{equation}
 \item $\theta_k(z,\tau)$ $(k=0,1,2)$ satisfy the following duplication formulas:
\begin{equation}
\begin{split}
& \theta_0(0,\tau)^3\theta_0(2z,\tau) = \theta_1(z,\tau)\left[\theta_2(z,\tau)^3-\theta_0(z,\tau)^3\right],\\
& \theta_0(0,\tau)^3\theta_1(2z,\tau) = \theta_0(z,\tau)\left[\theta_1(z,\tau)^3-\theta_2(z,\tau)^3\right],\\
& \theta_0(0,\tau)^3\theta_2(2z,\tau) = \theta_2(z,\tau)\left[\theta_0(z,\tau)^3-\theta_1(z,\tau)^3\right].
\end{split}\label{eqn:th_dup}
\end{equation}
 \end{enumerate}
\end{proposition}
It seems that the above formulas are well known\cite{BL:cubic}, but however, it might be useful for
non-experts to give an elementary proof here. In the following, we fix $\tau\in{\cal H}$ and write
$\theta_k(z,\tau)=\theta_k(z)$.
\begin{lemma}\label{lem:addition}
 $\theta_k(z,\tau)$ ($k=0,1,2$) satisfy the following addition formulas:
\begin{eqnarray}
&& \theta_0(0)^2\theta_0(x+y)\theta_0(x-y)=\theta_1(x)\theta_2(x)\theta_2(y)^2 - \theta_0(x)^2\theta_0(y)\theta_1(y),\label{eqn:th_ad1}\\
&& \theta_0(0)^2\theta_1(x+y)\theta_0(x-y) = \theta_0(x)\theta_1(x)\theta_1(y)^2-\theta_2(x)^2\theta_0(y)\theta_2(y),\label{eqn:th_ad2}\\
&& \theta_0(0)^2\theta_2(x+y)\theta_0(x-y) = \theta_0(x)\theta_2(x)\theta_0(y)^2-\theta_1(x)^2\theta_1(y)\theta_2(y),\label{eqn:th_ad3}\\
&& \theta_0(0)^2\theta_0(x+y)\theta_1(x-y) = \theta_0(x)\theta_1(x)\theta_0(y)^2-\theta_2(x)^2\theta_1(y)\theta_2(y),\label{eqn:th_ad4}\\
&& \theta_0(0)^2\theta_1(x+y)\theta_1(x-y)=\theta_0(x)\theta_2(x)\theta_2(y)^2-\theta_1(x)^2\theta_0(y)\theta_1(y),\label{eqn:th_ad5}\\
&& \theta_0(0)^2\theta_2(x+y)\theta_1(x-y)=\theta_1(x)\theta_2(x)\theta_1(y)^2-\theta_0(x)^2\theta_0(y)\theta_2(y),\label{eqn:th_ad6}\\
&& \theta_0(0)^2\theta_0(x+y)\theta_2(x-y)=\theta_0(x)\theta_2(x)\theta_1(y)^2-\theta_1(x)^2\theta_0(y)\theta_2(y),\label{eqn:th_ad7}\\
&&\theta_0(0)^2\theta_1(x+y)\theta_2(x-y) = \theta_1(x)\theta_2(x)\theta_0(y)^2-\theta_0(x)^2\theta_1(y)\theta_2(y),\label{eqn:th_ad8}\\
&&\theta_0(0)^2\theta_2(x+y)\theta_2(x-y) = \theta_0(x)\theta_1(x)\theta_2(y)^2-\theta_2(x)^2\theta_0(y)\theta_1(y).\label{eqn:th_ad9}
\end{eqnarray}
\end{lemma}
We give the proof of Lemma \ref{lem:addition} in the appendix.\par\medskip

\noindent{\bf Proof of Proposition \ref{prop:theta}.}\hfill\\ The duplication formulas
(\ref{eqn:th_dup}) are obtained by putting $x=y=z$ in eqs.~(\ref{eqn:th_ad1}), (\ref{eqn:th_ad2})
and (\ref{eqn:th_ad3}).  In order to prove eqs.~(\ref{eqn:th_Hesse}) and (\ref{eqn:elliptic_fn}), we
first note that it follows by definition that
\begin{equation}
 \theta_0(-z)=-\theta_1(z),\quad\theta_2(-z)=-\theta_2(z),\label{eqn:th_rel1}
\end{equation}
and hence
\begin{equation}
 \theta_1(0)=-\theta_0(0),\quad \theta_2(0)=0.\label{eqn:th_rel2}
\end{equation}
From Lemma \ref{lem:addition}, we obtain the addition formulas for
$\varphi(z)$ and $\psi(z)$ (see eq.~(\ref{eqn:elliptic_fn})) as
\begin{eqnarray}
 \varphi(x+y)&=& \frac{\varphi(x)\varphi(y)^2-\psi(x)^2\psi(y)}{\varphi(x)\psi(x)\psi(y)^2-\varphi(y)},\label{eqn:el_ad1}\\
 \varphi(x+y)&=& \frac{\psi(x)\psi(y)^2-\varphi(x)^2\varphi(y)}{\varphi(x)-\psi(x)^2\varphi(y)\psi(y)},\label{eqn:el_ad2}\\
 \varphi(x+y)&=& \frac{\varphi(x)\psi(x)-\varphi(y)\psi(y)}{\psi(x)\varphi(y)^2-\varphi(x)^2\psi(y)},\label{eqn:el_ad3}
\end{eqnarray}
\begin{eqnarray}
 \psi(x+y) &=& \frac{\psi(x)-\varphi(x)^2\varphi(y)\psi(y)}{\varphi(x)\psi(x)\psi(y)^2-\varphi(y)},\label{eqn:el_ad4}\\
 \psi(x+y) &=& \frac{\varphi(x)\psi(x)\varphi(y)^2-\psi(y)}{\varphi(x)-\psi(x)^2\varphi(y)\psi(y)},\label{eqn:el_ad5}\\
 \psi(x+y) &=& \frac{\varphi(x)\psi(y)^2-\psi(x)^2\varphi(y)}{\psi(x)\varphi(y)^2-\varphi(x)^2\psi(y)}.\label{eqn:el_ad6}
\end{eqnarray}
Differentiating eqs.~(\ref{eqn:el_ad1}) and (\ref{eqn:el_ad3}) by $y$ and putting $y=0$, we have
\begin{equation}
\varphi'(x)=-\varphi'(0)\varphi(x)-\psi'(0)\psi(x)^2,\quad
\varphi'(x) = \frac{\psi'(0)+2\varphi'(0)\varphi(x)\psi(x)+\psi'(0)\varphi(x)^3}{\psi(x)},\label{eqn:th9}
\end{equation}
respectively. Here we have used 
\begin{equation}
\varphi(0)=-1,\quad \psi(0)=0,
\end{equation}
which follows from eq.~(\ref{eqn:th_rel2}). Equating the right hand sides of
the two equations in eq.~(\ref{eqn:th9}), we have
\begin{equation}
1+\varphi(x)^3+\psi(x)^3=-3\frac{\varphi'(0)}{\psi'(0)}~\varphi(x)\psi(x),
\end{equation}
which yields eq.~(\ref{eqn:th_Hesse}) by multiplying $\theta_0(z)^3$. This completes the proof.\qed
\par\medskip
Consider the map
\begin{equation}
 \mathbb{C}\ni z\longmapsto [\theta_0(z):\theta_1(z):\theta_2(z)]\in\mathbb{P}^2(\mathbb{C}).
\end{equation}
From the relations
\begin{equation}
 \theta_k(z+1)=-\theta_k(z),\quad \theta_k(z+\tau)=-e^{3\pi i \tau-6\pi i z}\theta_k(z)\quad (k=0,1,2),
\end{equation}
we see that this induces a map from the complex torus
$L_\tau=\mathbb{C}/(\mathbb{Z}+\mathbb{Z}\tau)$ to $E_\mu$, which is known to give an isomorphism
$L_\tau\simeq E_\mu$ (see, e.g. \cite{BL:cubic}). Since $0\mapsto[1:-1:0]$, the addition formulas
(\ref{eqn:th_ad1})--(\ref{eqn:th_ad9}) induce the group structure on $E_\tau$ with the origin
$[1:-1:0]$. Denoting the addition of two points $[x_0:x_1:x_2]$ and $[x_0':x_1':x_2']$ as
$[x_0:x_1:x_2]\oplus[x_0':x_1':x_2']$, eqs.~(\ref{eqn:th_ad1})-(\ref{eqn:th_ad9}) imply
\begin{eqnarray}
&& [x_0:x_1:x_2]\oplus[x_0':x_1':x_2']\nonumber\\
&=&[x_1x_2x_2'^2-x_0^2x_0'x_1': x_0x_1x_1'^2-x_2^2x_0'x_2': x_0x_2x_0'^2-x_1^2x_1'x_2']\\
&=&[x_0x_1x_0'^2-x_2^2x_1'x_2': x_0x_2x_2'^2-x_1^2x_0'x_1': x_1x_2x_1'^2-x_0^2x_0'x_2']\\
&=&[x_0x_2x_1'^2-x_1^2x_0'x_2': x_1x_2x_0'^2-x_0^2x_1'x_2': x_0x_1x_2'^2-x_2^2x_0'x_1'].
\end{eqnarray}
In particular, when the two points are equal, the duplication formula is given by
\begin{equation}
2 [x_0:x_1:x_2]=[x_1(x_2^3-x_0^3): x_0(x_1^3-x_2^3): x_2(x_0^3-x_1^3)].
\end{equation}
Moreover, the inverse of $[x_0:x_1:x_2]$ is given by
\begin{equation}
 -[x_0:x_1:x_2]=[x_1:x_0:x_2].
\end{equation}

We finally remark that $\mu$ can be also expressed as follows. Differentiating
both equations in eq.~(\ref{eqn:elliptic_fn}) and putting $z=0$, we have by using eq.~(\ref{eqn:th_rel2})
\begin{equation}
 \varphi'(0)=2\frac{\theta_0'(0)}{\theta_0(0)},\quad \psi'(0)=\frac{\theta_2'(0)}{\theta_0(0)},
\end{equation}
which yield
\begin{equation}
 \mu(\tau)=-\frac{\varphi'(0)}{\psi'(0)}=-2\frac{\theta_0'(0)}{\theta_2'(0)}.\label{eqn:mu}
\end{equation}
\section{Ultradiscretization}
So far we have constructed the piecewise linear map (\ref{map:u-Hesse2}) as the duplication map on
the tropical cubic curve $C_K$, whose general solution is given by eq.~(\ref{gen_sol:u-Hesse}). We
have also presented the rational map (\ref{map:Hesse2}) which arises as the duplication map on the
Hesse cubic curve $E_\mu$. The general solution of the map is given by
eq.~(\ref{eqn:gen_sol:Hesse}).  In this section, we establish a correspondence between the two maps
and their general solutions by means of the ultradiscretization.
\subsection{Ultradiscretization of map}
The key of the ultradiscretization is the following formula:
\begin{equation}
 \lim_{\epsilon\to +0}\epsilon\log\left(e^{\frac{A}{\epsilon}}+e^{\frac{B}{\epsilon}}+\cdots\right)
=\max[A,B,\ldots].\label{eqn:formula_ultra}
\end{equation}
Putting
\begin{equation}
 x_n = e^{\frac{X_n}{\epsilon}},\quad  y_n = e^{\frac{Y_n}{\epsilon}},
\end{equation}
we have from eq.~(\ref{map:Hesse2})
\begin{equation}
\begin{split}
X_{n+1}& 
=\epsilon\log\left(1+e^{\frac{3X_n+\epsilon\pi i}{\epsilon}}\right) + Y_n 
- \epsilon\log\left(e^{\frac{3X_n}{\epsilon}} + e^{\frac{3Y_n+\epsilon\pi i}{\epsilon}}\right), \\
Y_{n+1}&
=\epsilon\log\left(1+e^{\frac{3Y_n+\epsilon\pi i}{\epsilon}}\right) + X_n 
- \epsilon\log\left(e^{\frac{3X_n}{\epsilon}} + e^{\frac{3Y_n+\epsilon\pi i}{\epsilon}}\right), 
\end{split}
\label{eqn:limit_process_map}
\end{equation}
which yields, in the limit $\epsilon\to +0$, eq.~(\ref{map:u-Hesse2}):
\begin{equation}
X_{n+1}=\max[0,3X_n]+Y_n-\max[3X_n,3Y_n],\quad
Y_{n+1}=\max[0,3Y_n]+X_n-\max[3X_n,3Y_n].
\end{equation}
The limit of the invariant curve (\ref{eqn:Hesse_curve}) yields $\overline{C}_K$:
\begin{equation}
 \max[0,3X,3Y]=X+Y+K,
\end{equation}
by the use of 
\begin{equation}
 3\mu(\tau)=e^{\frac{K}{\epsilon}}.\label{eqn:mu_K}
\end{equation}

In the above process of the ultradiscretization, we have calculated formally, for example, as
\begin{equation}
 \epsilon\log\left(1-e^{\frac{3X_n}{\epsilon}}\right)
= \epsilon\log\left(1+e^{\frac{3X_n+\epsilon\pi  i}{\epsilon}}\right)
\longrightarrow \max[0,3X]\quad (\epsilon\to +0).
\end{equation}
However, when the original rational map contains the minus signs, such formal calculation sometimes
does not give consistent result. This may happen, for example, when we consider the limit of the
exact solutions simultaneously, or when we consider the limit of the maps which are representation
of certain group or algebra. In both cases, the cancellations caused by the minus signs play a
crucial role on the level of rational maps, and the structure of the rational maps is lost because
such cancellations do not happen after taking the limit.  This problem is sometimes called the minus-sign
problem.

Therefore, we usually consider the subtraction-free rational map to apply the
ultradiscretization\cite{TT:tropicalWeyl,Tsuda:tropicalWeyl,Y:Amoebae}\footnote{It should be
remarked that the term ``tropical'' has been used differently in the communities of geometry and
integrable systems\cite{Kirillov}. In the former community it has been used to mean piecewise linear
objects, while in the latter subtraction-free rational maps. In the latter community the terms
``crystal'' or ``ultradiscrete'' have been used for piecewise linear objects. Therefore, it
sometimes happens that the term ``tropicalization'' can be used with opposite meanings.}, or we try
to transform the map to be subtraction-free if possible\cite{KNT:1d}. Unfortunately, it seems that
the map (\ref{map:Hesse2}) cannot be transformed to be subtraction-free by simple
transformations. However, in this case, it is possible to obtain valid ultradiscrete limit of the
general solution in spite of the minus-sign problem.
\begin{remark}\rm
The nine inflection points of the Hesse cubic curve correspond to the vertices of the tropical cubic
curve $C_K$ in the following manner; consider one of the inflection points
$[x_0:x_1:x_2]=[1:-1:0]=[e^{\frac{0}{\epsilon}}:e^{\frac{0+i\pi
\epsilon}{\epsilon}}:e^{\frac{-\infty}{\epsilon}}]$.  Then putting $x_i=e^{\frac{X_i}{\epsilon}}$
($i=0,1,2$) and taking the limit $\epsilon\to +0$, we have
$[X_0:X_1:X_2]=[0:0:-\infty]=[\infty:\infty:0]$. Note here that on this level equivalence of the
homogeneous coordinates is given by $[X_0:X_1:X_2]=[X_0+L:X_1+L:X_2+L]$ for any constant $L$. In the
inhomogeneous coordinates, this point corresponds to $(\infty,\infty)$, which is linearly equivalent
to the vertex $V_3=(K,K)$. Similarly, the two points $[1:-\omega:0],\ [1:-\omega^2:0]$ also
correspond to $V_3$. Furthermore, the triple of points $\{[1:0:-1],\ [1:0:-\omega],\
[1:0:-\omega^2]\}$ correspond to $V_2=(0,-K)$, and the triple $\{[0:1:-1],\ [0:1:-\omega],\
[0:1:-\omega^2]\}$ to $V_1=(-K,0)$. In other words, three inflection points of the Hesse cubic curve
degenerate to each vertex of $C_K$ in the ultradiscrete limit. This explains the reason why the
multiplicity of each vertex of $C_K$ is $3$ and $C_K$ is not smooth while the Hesse cubic curve is
non-singular.
\end{remark}
\subsection{Ultradiscretization of general solution}
In this section, we consider the ultradiscrete limit of the solution. The following is the main
result of this paper.
\begin{theorem} \label{theorem:main}
The general solution (\ref{eqn:gen_sol:Hesse}) of the rational map (\ref{map:Hesse2}) reduces to
the general solution (\ref{gen_sol:u-Hesse}) of the piecewise linear map (\ref{map:u-Hesse2}) by
 taking the limit $\epsilon\to +0$ under the parametrization 
\begin{equation}
 X_n = e^{\frac{x_n}{\epsilon}},\quad  Y_n = e^{\frac{y_n}{\epsilon}},\quad
  \frac{\tau}{\tau+\frac{1}{3}} = -\frac{9K}{2\pi i\epsilon},\quad
z_0 = \frac{u_0}{9K}\left(1-\frac{2\pi i \epsilon}{9K}\right),\quad u_0\in\mathbb{R},\quad K>0.\label{eqn:ultra_par}
\end{equation}
\end{theorem}
\par\bigskip
The ultradiscrete limit of the theta function can be realized by taking ${\rm Im}~\tau\to 0$,
however, the limit of the real part of $\tau$ should be carefully chosen in order to obtain
consistent result\cite{Nobe:utheta}. For choosing the limit of real part of $\tau$, the following
observation on the correspondence between the zeros of the theta functions and non-smooth points of
$S(u;\alpha,\beta,\theta)$ is crucial. \par\medskip

\begin{figure}[h]
 \hfil \includegraphics[height=4cm]{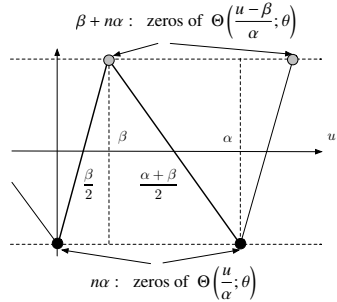}\hfil
\caption{Zigzag pattern of
 $S(u;\alpha,\beta,\theta)=\Theta(\frac{u}{\alpha};\theta)-\Theta(\frac{u-\beta}{\alpha};\theta)$
 and the zeros of $\Theta(\frac{u}{\alpha};\theta)$, $\Theta(\frac{u-\beta}{\alpha};\theta)$.}\label{fig:s2}
\end{figure}

\begin{figure}[h]
 \hfil \includegraphics[height=4cm]{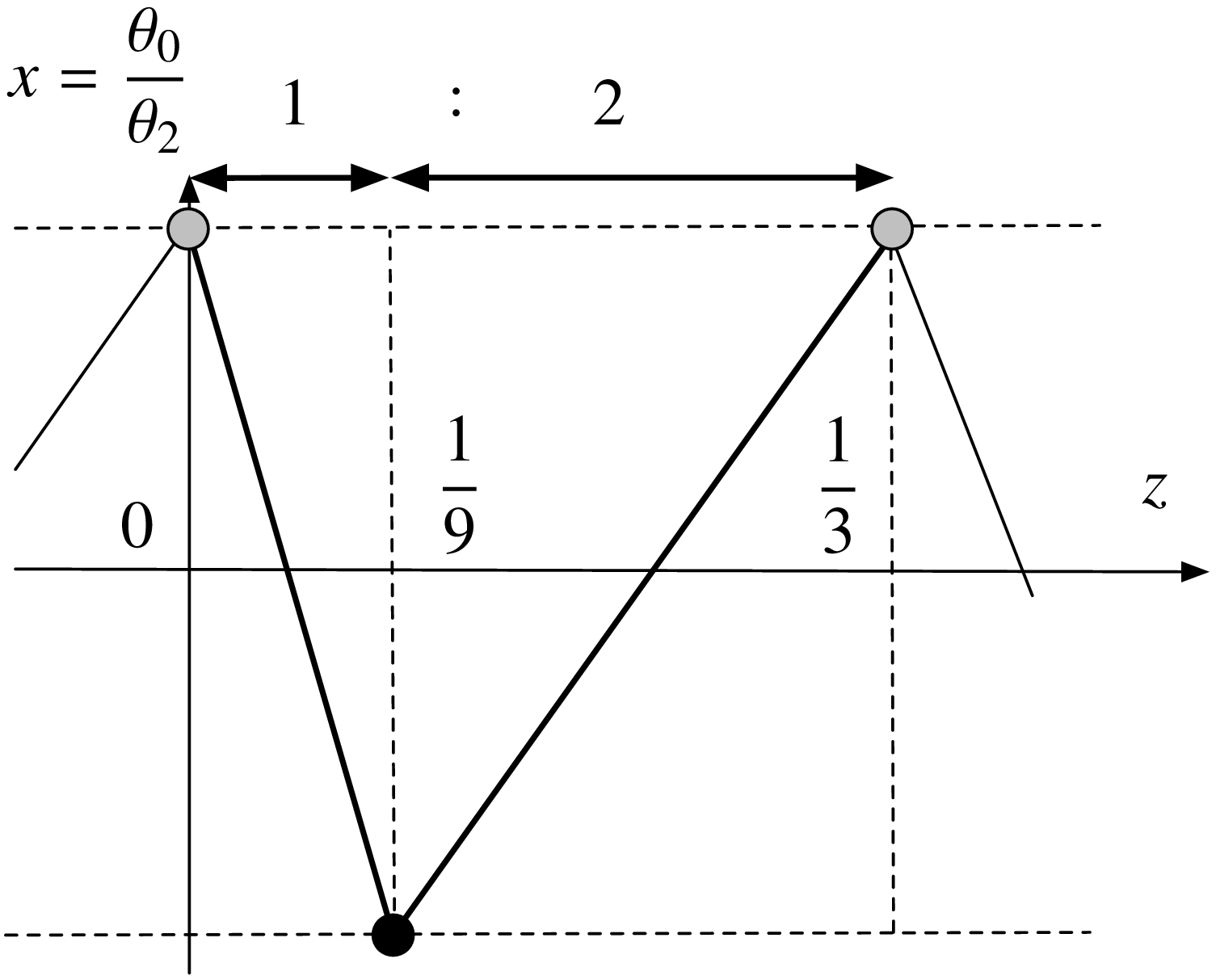}\qquad
\includegraphics[height=4cm]{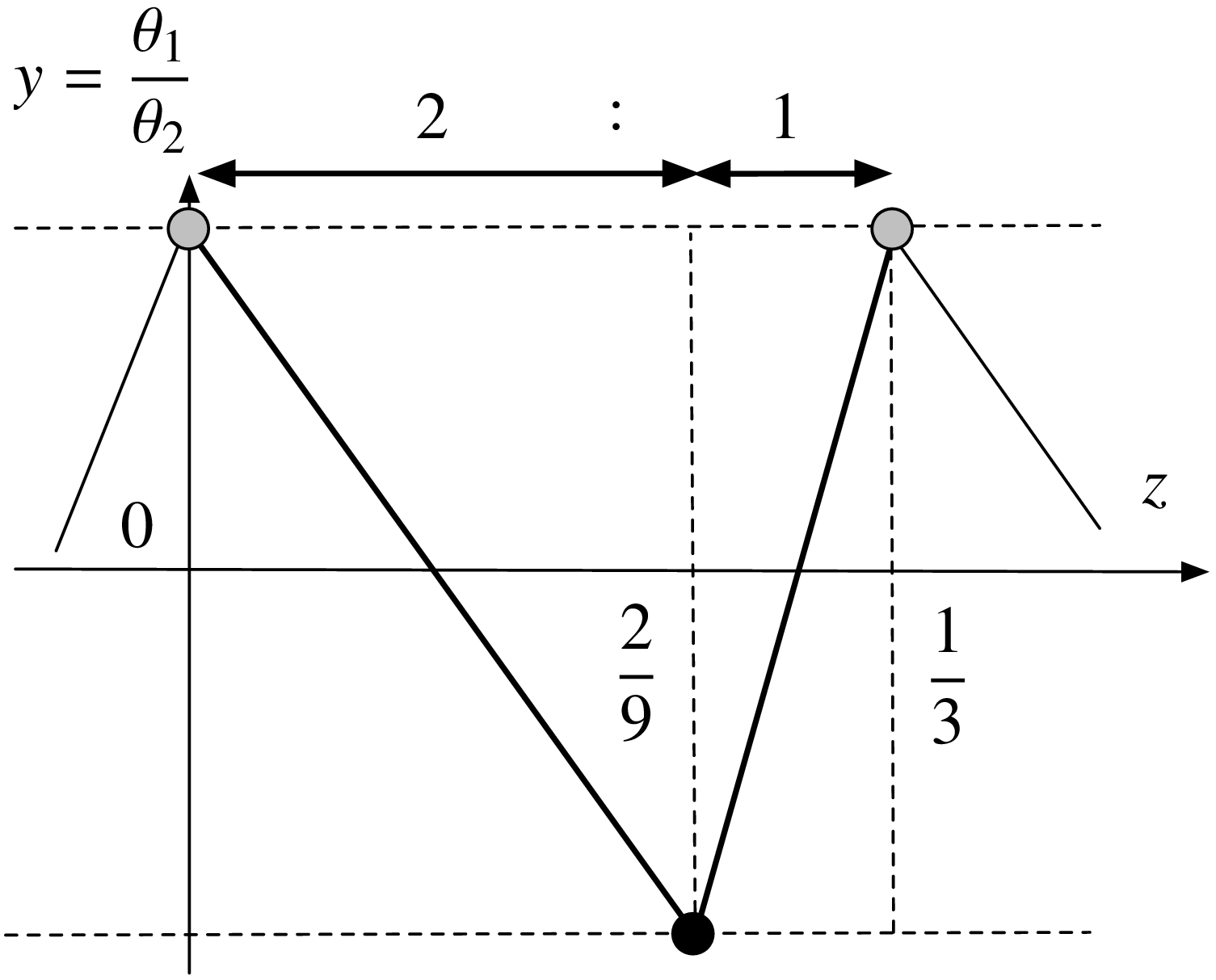}\hfil
\caption{Zigzag patterns obtained by the limit $\tau\to-\frac{1}{3}$.}\label{fig:patterns}
\end{figure}

\noindent\textbf{Observation:}\quad The ultradiscrete theta function $\Theta(u;\theta)$ defined by
eq.~(\ref{eqn:u-theta}) is a piecewise quadratic function with the period $1$, and has zeros at
$u=n\in\mathbb{Z}$.  $\Theta(u;\theta)$ can be obtained from
$\vartheta_0(z;\tau)=\vartheta_{(0,\frac{1}{2})}(z;\tau)$ by taking the limit $\tau\to
0$\cite{Nobe:utheta}. Since the zeros of $\vartheta_0(z;\tau)$ are located at
$z=(m+\frac{1}{2})\tau+n$ ($m,n\in\mathbb{Z}$), the real zeros of $\vartheta_0(z;\tau)$ survive
under the limit, giving the zeros of $\Theta(u;\theta)$ at $u=n$. From the definition of $S$ given
in eq.~(\ref{eqn:S}) and Fig.\ref{fig:S}, it is easy to see that the valleys at $u=n\alpha$ and the
peaks at $u=\beta+n\alpha$ ($n\in\mathbb{Z}$) of $S$ correspond to the zeros of
$\Theta(\frac{u}{\alpha};\theta)$ and $\Theta(\frac{u-\beta}{\alpha};\theta)$, respectively, as
illustrated in Fig.\ref{fig:s2}. In other words, valleys and peaks of the ultradiscrete elliptic
function $S$ arise from the zeros and poles of the corresponding elliptic function, respectively.
Now, noticing that the zeros of $\vartheta_{(a,b)}(z,\tau)$ are located at $z=(-a+m+\frac{1}{2})\tau + (-b+n+\frac{1}{2})$
$(m,n\in\mathbb{Z}$), the zeros of $\theta_k(z,\tau)$ ($k=0,1,2$) are given by
\begin{eqnarray*}
&& \theta_0(z,\tau):\quad z = \left(m+\frac{2}{3}\right)\tau+\frac{1}{3}(n-1),\\
&& \theta_1(z,\tau):\quad z = \left(m+\frac{1}{3}\right)\tau+\frac{1}{3}(n-1),\\
&& \theta_2(z,\tau):\quad z = m\tau+\frac{1}{3}(n-1),
\end{eqnarray*}
respectively. It is obvious that the zeros and poles of
$x_n=\frac{\theta_0(z,\tau)}{\theta_2(z,\tau)}$ and $y_n=\frac{\theta_1(z,\tau)}{\theta_2(z,\tau)}$
cancel each other, respectively, in the limit $\tau\to 0$, which yields trivial result. Let us choose
$\tau\to-\frac{1}{3}$. Then the zeros of $\theta_i(z,\tau)$ ($i=0,1,2$) become
\begin{eqnarray*}
&& \theta_0(z,\tau):\quad z =\frac{\mathbb{Z}}{3}-\frac{2}{9} = \frac{\mathbb{Z}}{3}+\frac{1}{9},\\
&& \theta_1(z,\tau):\quad z = \frac{\mathbb{Z}}{3}-\frac{1}{9} = \frac{\mathbb{Z}}{3}+\frac{2}{9},\\
&& \theta_2(z,\tau):\quad z = \frac{\mathbb{Z}}{3},
\end{eqnarray*}
respectively, which give the zigzag patterns in the limit as illustrated in
Fig.\ref{fig:patterns}. These patterns would coincide with the those in Fig.\ref{fig:projections}
after an appropriate scaling.

Before proceeding to the proof of the Theorem \ref{theorem:main}, we prepare the modular
transformation of the theta function, which is useful in taking the limit of $\tau$.
\begin{proposition}\cite{Igusa,Mumford}\label{prop:modular}
\begin{equation}
 \vartheta_{\sigma\cdot m}(\sigma\cdot z,\sigma\cdot\tau)
 =e^{\pi i(\sigma\cdot z)cz}~(c\tau+d)^{\frac{1}{2}}~\kappa(\sigma)~e^{2\pi i \phi_m(\sigma)}~
\vartheta_{m}(z,\tau),
\end{equation}
where
 \begin{eqnarray}
&&\left\{
\begin{array}{l}
{\displaystyle m=(m_1,m_2),\quad
\sigma=\left(\begin{array}{cc} a&b \\c&d\end{array}\right) \in{\rm SL}_2(\mathbb{Z})}, \\[4mm]
{\displaystyle \sigma\cdot m=m\sigma^{-1}+\frac{1}{2}(cd,ab)},\quad
{\displaystyle \sigma\cdot\tau = \frac{a\tau+b}{c\tau+d}},\quad 
{\displaystyle \sigma\cdot z = \frac{z}{c\tau+d}},\\[4mm]
{\displaystyle
\phi_m(\sigma)=-\frac{1}{2}\left[bdm_1^2 + acm_2^2 - 2bcm_1m_2-ab(dm_1-cm_2)\right]},\\[4mm]
{{\displaystyle \kappa(\sigma)}}:\ \mbox{an eighth root of 1 depending only on $\sigma$.}
\end{array}\right.
\end{eqnarray}
\end{proposition}
\begin{remark}\rm
 Explicit expression of $\kappa(\sigma)$ is given by \cite{Ibukiyama}
\begin{displaymath}
 \kappa(\sigma) = \left\{
\begin{array}{ll}
 {\displaystyle e^{\pi i
  (\frac{abcd}{2}+\frac{acd^2}{4}-\frac{c}{4})}~\left(\frac{a}{|c|}\right)},& c\ :~\mbox{odd}\\
{\displaystyle
 \left(\frac{c}{|d|}\right)(-1)^{({\rm sgn}(c)-1)({\rm sgn}(d)-1)/4}~e^{\frac{\pi i}{4}(d-1)}},& c:~\mbox{even}
\end{array}
\right.
\end{displaymath}
where $\left(\frac{a}{p}\right)$ is the Legendre (Jacobi) symbol for the quadratic residue. In particular, $\left(\frac{a}{1}\right)=1$.
\end{remark}
\textbf{Proof of Theorem \ref{theorem:main}.}\quad The first key of the proof is to 
apply the modular transformation on $\theta_k(z,\tau)=\vartheta_{\left(\frac{k}{3}-\frac{1}{6},\frac{3}{2}\right)}(3z,3\tau)$
($k=0,1,2$) specified by
\begin{equation}
 \sigma=\left(\begin{array}{cc}1 &0 \\1 & 1\end{array}\right).\label{eqn:sigma}
\end{equation}
From 
\begin{equation}
\sigma\cdot m =\left( \frac{k}{3}-\frac{7}{6},\frac{3}{2}\right),\quad
\phi_m(\sigma) = -  \frac{9}{8},\quad \kappa(\sigma) = 1,\label{modular_transf:data1}
\end{equation}
we have
\begin{eqnarray}
\theta_k(z,\tau)=\vartheta_{(\frac{k}{3}-\frac{1}{6},\frac{3}{2})}(3z,3\tau)=
e^{-\frac{3\pi i z^2}{\tau+\frac{1}{3}}}~(3\tau+1)^{-\frac{1}{2}}~
e^{\frac{1}{4}\pi i}~ \vartheta_{(\frac{k}{3}-\frac{7}{6},\frac{3}{2} )}
\left(\frac{z}{\tau+\frac{1}{3}},\frac{\tau}{\tau+\frac{1}{3}}\right).
\label{theta:transformed}
\end{eqnarray}
We put
\begin{equation}
 \frac{\tau}{\tau+\frac{1}{3}} = -\frac{\theta}{i\pi \epsilon},\quad \theta>0,\label{modular_transf:tau}
\end{equation}
and take the limit of $\epsilon\to +0$ which corresponds to $\tau\to -\frac{1}{3}$. Noticing
$ \frac{z}{\tau+\frac{1}{3}}=3\left(\frac{\theta}{i\pi\epsilon}+1\right)z$, we have
\begin{equation}
\vartheta_{(\frac{k}{3}-\frac{7}{6},\frac{3}{2})}
\left(\frac{z}{\tau+\frac{1}{3}},\frac{\tau}{\tau+\frac{1}{3}}\right)
=\sum_{n\in\mathbb{Z}}e^{-\frac{\theta}{\epsilon}(n+\frac{k}{3}-\frac{7}{6})^2 + \frac{6\theta}{\epsilon}(n+\frac{k}{3}-\frac{7}{6})z}
~ e^{6\pi i(n+\frac{k}{3}-\frac{7}{6}) \left(z+\frac{1}{2}\right)}.\label{eqn:theta_ul1}
\end{equation}

The second key is to use the freedom of the imaginary part of $z\in\mathbb{C}$. Putting
\begin{equation}
  z= \frac{u}{9K} + iv,\quad K,\ u,\ v\in\mathbb{R},\label{eqn:z}
\end{equation}
eq.~(\ref{eqn:theta_ul1}) is rewritten as
\begin{eqnarray}
&& \vartheta_{(\frac{k}{3}-\frac{7}{6},\frac{3}{2})}
\left(\frac{z}{\tau+\frac{1}{3}},\frac{\tau}{\tau+\frac{1}{3}}\right)
=\sum_{n\in\mathbb{Z}}e^{-\frac{\theta}{\epsilon}(n+\frac{k}{3}-\frac{7}{6})^2 
+ \frac{6\theta}{\epsilon}(n+\frac{k}{3}-\frac{7}{6})(\frac{u}{9K} + iv)}
~ e^{6\pi i(n+\frac{k}{3}-\frac{7}{6}) \left(\frac{u}{9K} + iv + \frac{1}{2}\right)}\nonumber\\
&=&
e^{\frac{\theta }{9K^2\epsilon}u^2}
\sum_{n\in\mathbb{Z}}e^{-\frac{\theta}{\epsilon}\left[\frac{u}{3K}-n-\frac{k}{3}+\frac{7}{6}\right]^2 
- 6\pi (n+\frac{k}{3}-\frac{7}{6})v}
~ e^{6\pi i(n+\frac{k}{3}-\frac{7}{6})\left[\frac{\theta}{\pi\epsilon}v+  \frac{u}{9K} + \frac{1}{2}\right] }.
\label{eqn:theta_ul2}
\end{eqnarray}
If $u$ and $v$ satisfy
\begin{equation}
 \frac{\theta}{\pi\epsilon}v+ \frac{u}{9K}=0,\label{eqn:uv}
\end{equation}
then eq.~(\ref{eqn:theta_ul2}) is simplified as
\begin{eqnarray*}
 \vartheta_{(\frac{k}{3}-\frac{7}{6},\frac{3}{2})}
\left(\frac{z}{\tau+\frac{1}{3}},\frac{\tau}{\tau+\frac{1}{3}}\right)
&=& e^{\frac{\theta }{9K^2\epsilon}u^2}~e^{\pi i(k-\frac{7}{2})}
\sum_{n\in\mathbb{Z}}e^{-\frac{\theta}{\epsilon}\left[\frac{u}{3K}-n-\frac{k}{3}+\frac{7}{6}\right]^2 
- 6\pi (n+\frac{k}{3}-\frac{7}{6})v}
~ e^{3\pi in}\\
&=&
 e^{\frac{\theta }{9K^2\epsilon}u^2}~e^{\pi i(k-\frac{7}{2})}
\sum_{n\in\mathbb{Z}}e^{-\frac{\theta}{\epsilon}\left[\frac{u-K(k+1)}{3K}-(n-2)-\frac{1}{2}\right]^2 
+ (n+\frac{k}{3}-\frac{7}{6})\frac{2\pi^2 \epsilon}{3K\theta} u}
~ e^{3\pi in}.
\end{eqnarray*}
Therefore, asymptotic behaviors of $x_n$ and $y_n$ as $\epsilon~\sim +0$ are given by 
\begin{equation}
\begin{split}
 x_n=&\frac{\theta_0(z,\tau)}{\theta_2(z,\tau)}
\sim\frac{
\sum\limits_{n\in\mathbb{Z}} 
e^{-\frac{\theta}{\epsilon}
\left[\frac{u-K}{3K}-(n-2)-\frac{1}{2} \right]^2}
e^{3 n \pi i}
}
{
e^{2\pi i}
\sum\limits_{n\in\mathbb{Z}} 
e^{-\frac{\theta}{\epsilon}
\left[\frac{u-3K}{3K}-\frac{1}{2}-(n-2)\right]^2}e^{3n\pi i}}
\sim
e^{(n_0+n_2)\pi i}
\frac{
e^{-\frac{\theta}{\epsilon}
\left[\left(\left( \frac{u-K}{3K}\right)\right)-\frac{1}{2} \right]^2}
}
{
e^{-\frac{\theta}{\epsilon}
\left[\left(\left(\frac{u-3K}{3K}\right)\right)-\frac{1}{2} \right]^2}
},\\[4mm]
 y_n=&\frac{\theta_1(z,\tau)}{\theta_2(z,\tau)}
\sim\frac{
e^{\pi i}
\sum\limits_{n\in\mathbb{Z}} 
e^{-\frac{\theta}{\epsilon}
\left[\frac{u-2K}{3K}-(n-2)-\frac{1}{2} \right]^2}
e^{3 n \pi i}
}
{
e^{2\pi i}
\sum\limits_{n\in\mathbb{Z}} 
e^{-\frac{\theta}{\epsilon}
\left[\frac{u-3K}{3K}-(n-2)-\frac{1}{2} \right]^2}e^{3n\pi i}}
\sim
e^{(n_1+n_2+1)\pi i}
\frac{
e^{-\frac{\theta}{\epsilon}
\left[\left(\left(\frac{u-2K}{3K} \right)\right)-\frac{1}{2} \right]^2}
}
{
e^{-\frac{\theta}{\epsilon}
\left[\left(\left(\frac{u-3K}{3K}\right)\right)-\frac{1}{2} \right]^2}
},
\end{split}\label{eqn:theta_ul3}
\end{equation}
respectively, where 
\begin{equation}
n_0 = {\rm Floor}\left[\frac{u-K}{3K}\right]+2,
\quad n_1 = {\rm Floor}\left[\frac{u-2K}{3K}\right]+2,
\quad n_2 = {\rm Floor}\left[\frac{u-3K}{3K}\right]+2.
\end{equation}
Here, we note that $z$ and $u$ are understood as $2^nz_0$ and $2^nu_0$, respectively.

In order to obtain the final result, the remaining task is to relate the parameters $\theta$ with $K$ in the
limit $\epsilon\to +0$. This can be done by considering the limit of the conserved quantity
$\mu(\tau)$ given by eq.~(\ref{eqn:th_Hesse}) or eq.~(\ref{eqn:mu}). Noticing eq.~(\ref{eqn:mu}), we
differentiate $\theta_{0}(z,\tau)$ and $\theta_{2}(z,\tau)$ by $z$ after applying the modular
transformation. Then $\theta'_0(0,\tau)$ and $\theta'_1(0,\tau)$ can be calculated by using
eqs.~(\ref{theta:transformed}) and (\ref{eqn:theta_ul1}) as
\begin{eqnarray*}
&& \theta_0'(0,\tau)=\left.\frac{\partial}{\partial z}\vartheta_{(-\frac{1}{6},\frac{3}{2})}(3z,3\tau)\right|_{z=0}\\
&=&\left.\frac{\partial}{\partial z}
e^{-\pi i \left(1+\frac{\theta}{i\pi \epsilon}\right)z^2}\left(1+\frac{\theta}{i\pi \epsilon}\right)^{\frac{1}{2}}~e^{\frac{1}{4}\pi i}~
\sum_{n\in\mathbb{Z}}e^{-\frac{\theta}{\epsilon}(n-\frac{7}{6})^2 + \frac{6\theta}{\epsilon}(n-\frac{7}{6})z}
~ e^{6\pi i(n-\frac{7}{6}) \left(z+\frac{1}{2}\right)}
\right|_{z=0}
\\
&=&\left(1+\frac{\theta}{i\pi \epsilon}\right)^{\frac{1}{2}}~e^{\frac{3}{4}\pi i}~
\sum_{n\in\mathbb{Z}} \left[6\left( \pi i+\frac{\theta}{\epsilon}\right)
\left(n-\frac{7}{6}\right)\right]e^{-\frac{\theta}{\epsilon}\left(n-\frac{7}{6}\right)^2
+3n\pi i},
\end{eqnarray*}
\begin{eqnarray*}
&& \theta_2'(0,\tau)=\left.\frac{\partial}{\partial z}\vartheta_{(\frac{1}{2},\frac{3}{2})}(3z,3\tau)\right|_{z=0}\\
&=&\left.\frac{\partial}{\partial z}
e^{-\pi i \left(1+\frac{\theta}{i\pi \epsilon}\right)z^2}\left(1+\frac{\theta}{i\pi \epsilon}\right)^{\frac{1}{2}}~e^{\frac{1}{4}\pi i}~
\sum_{n\in\mathbb{Z}}e^{-\frac{\theta}{\epsilon}(n-\frac{1}{2})^2 + \frac{6\theta}{\epsilon}(n-\frac{1}{2})z}
~ e^{6\pi i(n-\frac{1}{2}) \left(z+\frac{1}{2}\right)}\right|_{z=0}
\\
&=&\left(1+\frac{\theta}{i\pi \epsilon}\right)^{\frac{1}{2}}~e^{\frac{3}{4}\pi i}~
\sum_{n\in\mathbb{Z}} \left[6\left( \pi i+\frac{\theta}{\epsilon}\right)
\left(n-\frac{1}{2}\right)\right]e^{-\frac{\theta}{\epsilon}\left(n-\frac{1}{2}\right)^2+3n\pi i},
\end{eqnarray*}
respectively, which imply
\begin{displaymath}
  3\mu
=-6~\frac{\theta_0'(0,\tau)}{\theta_2'(0,\tau)}
=-6\frac{
e^{\frac{3}{4}\pi i}~
\sum\limits_{n\in\mathbb{Z}} 
\left(n-\frac{7}{6}\right)e^{-\frac{\theta}{\epsilon}\left(n-\frac{7}{6}\right)^2 + 3n\pi i}
}
{e^{\frac{3}{4}\pi i}
\sum\limits_{n\in\mathbb{Z}}\left(n-\frac{1}{2}\right)
e^{-\frac{\theta}{\epsilon}\left(n-\frac{1}{2}\right)^2+3n\pi i}
}
\sim
-6~\frac{\left(-\frac{1}{6}\right)e^{-\frac{\theta}{36\epsilon}+3\pi i}}
{-\frac{1}{2}e^{-\frac{\theta}{4\epsilon}} + \frac{1}{2}e^{-\frac{\theta}{4\epsilon}+3\pi i}
}
=
e^{\frac{2\theta}{9\epsilon}},
\end{displaymath}
as $\epsilon\to +0$. Accordingly, from eq.~(\ref{eqn:mu_K}) we may put consistently as
\begin{equation}
 \theta=\frac{9}{2}K.\label{eqn:theta_K}
\end{equation}
Let us set $x_n=e^{\frac{X_n}{\epsilon}}$ and $y_n=e^{\frac{Y_n}{\epsilon}}$ in eq.~(\ref{eqn:theta_ul3}).
Then the complex factors in eq.~(\ref{eqn:theta_ul3}) disappear in the limit of $\epsilon\to 0$
and we finally obtain by using eq.~(\ref{eqn:theta_K})
\begin{equation}
\begin{split}
 &X_n=
-\frac{9K}{2}\left[\left(\left( \frac{u-K}{3K}\right)\right)-\frac{1}{2} \right]^2
+\frac{9K}{2}
\left[\left(\left(\frac{u-3K}{3K}\right)\right)-\frac{1}{2} \right]^2,\\
&Y_n=
-\frac{9K}{2}\left[\left(\left( \frac{u-2K}{3K}\right)\right)-\frac{1}{2} \right]^2
+\frac{9K}{2}
\left[\left(\left(\frac{u-3K}{3K}\right)\right)-\frac{1}{2} \right]^2,
\end{split}
\end{equation}
which is equivalent to eq.~(\ref{gen_sol:u-Hesse}). This completes the proof.\qed\par\bigskip

We finally remark that the choice of parametrization (\ref{eqn:ultra_par}) can be also justified by
the following observation. The asymptotic formula (\ref{eqn:theta_ul3}) shows that $(x_n,y_n)$ is
in $\mathbb{R}^2$ and that the quadrant of $(x_n,y_n)$ changes according to the value of
$((\frac{u}{3K}))$ as described in Table \ref{table:asymptotics}. Note that $(x_n,y_n)$ never enters
the first quadrant. It implies that qualitative behavior of the real orbit of the map
(\ref{map:u-Hesse2}) discussed in Section 3 is preserved under the limiting process.
\begin{table}[h]
\begin{center}
\begingroup
\renewcommand{\arraystretch}{1.5}
 \begin{tabular}{|c|ccc|cc||c|}
\hline
  $((\frac{u}{3K}))$        & $n_0$ & $n_1$&$n_2$ &$n_0+n_2$ &$n_1+n_2+1$ & $(x_n,y_n)$ \\
\hline
$[\frac{2}{3},1)$           & $N$   & $N$  &$N-1$ &$2N-1$    & $2N$       & $(-,+)$\\
\hline
$[\frac{1}{3},\frac{2}{3})$ & $N$   & $N-1$&$N-1$ &$2N-1$    & $2N-1$     & $(-,-)$\\
\hline
$[0,\frac{1}{3})$           & $N-1$ & $N-1$&$N-1$ &$2N-2$    & $2N-1$     & $(+,-)$\\
\hline
 \end{tabular} 
\caption{Quadrant of $(x_n,y_n)$ for $\epsilon\sim +0$, where $N={\rm Floor}~((\frac{u}{3K}))+2$.}\label{table:asymptotics}
\endgroup
\end{center}
\end{table}

\noindent\textbf{Acknowledgement}\quad The authors would like to express their sincere thanks to
Prof. A.~Nakayashiki and Prof.~Y. Yamada for fruitful discussions. This work was partially supported
by the JSPS Grant-in-Aid for Scientific Research No.~19340039, 19740086 and 19840039.

\appendix
\makeatletter
\@addtoreset{equation}{section}
\renewcommand{\theequation}{\@Alph\c@section.\@arabic\c@equation}
\makeatother
\section{Proof of Lemma \ref{lem:addition}}
In this appendix, we give a proof of Lemma \ref{lem:addition}.  Besides the cases $k=0,1,2$ we 
also use the cases $k=\frac{1}{2},\frac{3}{2},\frac{5}{2}$ as well. Note that
$\theta_{k+3}(z,\tau)=\theta_k(z,\tau)$.

First, we remark that $\theta_{\frac{1}{2}}(z,\tau)=\vartheta_{(0,\frac{3}{2})}(3z,3\tau)$ and
$\theta_{2}(z,\tau)=\vartheta_{(\frac{1}{2},\frac{3}{2})}(3z,3\tau)$ are $\vartheta(3z,3\tau)$ and
$\vartheta_1(3z,3\tau)$ of Jacobi's notation, respectively. Let us start from the eq.~(A)-(4) in
\cite{Jacobi}:
\begin{equation}
\begin{split}
& \theta_{\frac{1}{2}}(w) \theta_{\frac{1}{2}}(x) \theta_{\frac{1}{2}}(y) \theta_{\frac{1}{2}}(z)
- \theta_{2}(w) \theta_{2}(x) \theta_{2}(y) \theta_{2}(z)\\
&=  \theta_{\frac{1}{2}}(w') \theta_{\frac{1}{2}}(x') \theta_{\frac{1}{2}}(y') \theta_{\frac{1}{2}}(z')
- \theta_{2}(w') \theta_{2}(x') \theta_{2}(y') \theta_{2}(z'),\label{eqn:th1} 
\end{split}
\end{equation}
where
\begin{equation}
 w'=\frac{w+x+y+z}{2},\quad  x'=\frac{w+x-y-z}{2},\quad  y'=\frac{w-x+y-z}{2}, \quad z'=\frac{w-x-y+z}{2}.
\end{equation}
Replacing $w$ as $w\rightarrow w+\frac{\tau}{3}$, we have $w'\rightarrow w'+\frac{\tau}{6}$, 
$x'\rightarrow x'+\frac{\tau}{6}$, $y'\rightarrow y'+\frac{\tau}{6}$ and $z'\rightarrow
z'+\frac{\tau}{6}$. By using the formula
\begin{equation}
 \theta_k\left(z+\frac{\tau}{6}\right)=e^{-\frac{\pi i}{2}}~e^{-\frac{\pi i\tau}{12}-\pi i z}~\theta_{k+\frac{1}{2}}(z),\quad
 \theta_k\left(z+\frac{\tau}{3}\right)=e^{-\pi i}~e^{-\frac{\pi i\tau}{3}-2\pi i z}~\theta_{k+1}(z),\quad
\end{equation}
which easily follows by definition, we obtain
\begin{equation}
\begin{split}
& -\theta_{\frac{3}{2}}(w)\theta_{\frac{1}{2}}(x)\theta_{\frac{1}{2}}(y)\theta_{\frac{1}{2}}(z)
+ \theta_{0}(w)\theta_{2}(x)\theta_{2}(y)\theta_{2}(z)\\
&= \theta_{1}(w')\theta_{1}(x')\theta_{1}(y')\theta_{1}(z')
-\theta_{\frac{5}{2}}(w')\theta_{\frac{5}{2}}(x')\theta_{\frac{5}{2}}(y')\theta_{\frac{5}{2}}(z').\label{eqn:th2} 
\end{split}
\end{equation}
Replacing further as $w\rightarrow w+\frac{\tau}{3}$, $x\rightarrow x+\frac{\tau}{3}$, $y\rightarrow
y+\frac{\tau}{3}$ and $z\rightarrow z+\frac{\tau}{3}$, we see that  $w'\rightarrow w'+\frac{2\tau}{3}$ and
$x'$, $y'$, $z'$ are unchanged. Then we obtain
\begin{equation}
\begin{split}
& -\theta_{\frac{5}{2}}(w)\theta_{\frac{3}{2}}(x)\theta_{\frac{3}{2}}(y)\theta_{\frac{3}{2}}(z)
+ \theta_{1}(w)\theta_{0}(x)\theta_{0}(y)\theta_{0}(z)\\
&= \theta_{0}(w')\theta_{1}(x')\theta_{1}(y')\theta_{1}(z')
-\theta_{\frac{3}{2}}(w')\theta_{\frac{5}{2}}(x')\theta_{\frac{5}{2}}(y')\theta_{\frac{5}{2}}(z').\label{eqn:th3} 
\end{split}
\end{equation}
Application of the same transformation to eq.~(\ref{eqn:th1}) yields
\begin{equation}
\begin{split}
&  \theta_{\frac{3}{2}}(w)\theta_{\frac{3}{2}}(x)\theta_{\frac{3}{2}}(y)\theta_{\frac{3}{2}}(z)
- \theta_{0}(w)\theta_{0}(x)\theta_{0}(y)\theta_{0}(z)\\
&= \theta_{\frac{5}{2}}(w')\theta_{\frac{1}{2}}(x')\theta_{\frac{1}{2}}(y')\theta_{\frac{1}{2}}(z')
-\theta_{1}(w')\theta_{2}(x')\theta_{2}(y')\theta_{2}(z').\label{eqn:th4}
 \end{split}
\end{equation}

We put $w=-(x+y+z)$ in eqs.~(\ref{eqn:th2}) and (\ref{eqn:th3}). Then $w'=0$, $x'=-(y+z)$,
$y'=-(z+x)$ and $z'=-(x+y)$. By definition it follows that 
\begin{equation}
 \theta_0(-z)=-\theta_1(z),\quad \theta_{\frac{1}{2}}(-z)=\theta_{\frac{1}{2}}(z),\quad
\theta_2(-z)=-\theta_2(z),\quad \theta_{\frac{3}{2}}(-z)=\theta_{\frac{5}{2}}(z),\label{eqn:theta_rel1}
\end{equation}
and hence
\begin{equation}
 \theta_1(0)=-\theta_0(0),\quad \theta_{\frac{5}{2}}(0)=\theta_{\frac{3}{2}}(0),\quad \theta_2(0)=0.\label{eqn:theta_rel2}
\end{equation}
Therefore eqs.~(\ref{eqn:th2}) and (\ref{eqn:th3}) yield
\begin{equation}
\begin{split}
& -\theta_{\frac{5}{2}}(x+y+z)\theta_{\frac{1}{2}}(x)\theta_{\frac{1}{2}}(y)\theta_{\frac{1}{2}}(z)
- \theta_{1}(x+y+z)\theta_{2}(x)\theta_{2}(y)\theta_{2}(z)\\
&= \theta_{1}(0)\theta_{0}(y+z)\theta_{0}(z+x)\theta_{0}(x+y)
-\theta_{\frac{3}{2}}(0)\theta_{\frac{3}{2}}(y+z)\theta_{\frac{3}{2}}(z+x)\theta_{\frac{3}{2}}(x+y),\label{eqn:th5}
\end{split}
\end{equation}
\begin{equation}
\begin{split}
&  \theta_{\frac{3}{2}}(x+y+z)\theta_{\frac{3}{2}}(x)\theta_{\frac{3}{2}}(y)\theta_{\frac{3}{2}}(z)
+ \theta_{0}(x+y+z)\theta_{0}(x)\theta_{0}(y)\theta_{0}(z)\\
&= -\theta_{0}(0)\theta_{0}(y+z)\theta_{0}(z+x)\theta_{0}(x+y)
-\theta_{\frac{3}{2}}(0)\theta_{\frac{3}{2}}(y+z)\theta_{\frac{3}{2}}(z+x)\theta_{\frac{3}{2}}(x+y),\label{eqn:th6} 
\end{split}
\end{equation}
respectively. Similarly, putting $w=x+y+z$, we have that $w'=x+y+z$ and $x'$, $y'$, $z'$ are
unchanged. Then eq.~(\ref{eqn:th4}) yields 
\begin{equation}
\begin{split}
&  \theta_{\frac{3}{2}}(x+y+z)\theta_{\frac{3}{2}}(x)\theta_{\frac{3}{2}}(y)\theta_{\frac{3}{2}}(z)
- \theta_{0}(x+y+z)\theta_{0}(x)\theta_{0}(y)\theta_{0}(z)\\
& = \theta_{\frac{5}{2}}(x+y+z)\theta_{\frac{1}{2}}(x)\theta_{\frac{1}{2}}(y)\theta_{\frac{1}{2}}(z)
-\theta_{1}(x+y+z)\theta_{2}(x)\theta_{2}(y)\theta_{2}(z).\label{eqn:th7}
\end{split}
\end{equation}
Then from $-$(\ref{eqn:th5})+(\ref{eqn:th6})+(\ref{eqn:th7}) and dividing it by $2$, we have
\begin{equation}
 \theta_0(0)\theta_0(y+z)\theta_0(z+x)\theta_0(x+y)
=\theta_0(x+y+z)\theta_0(x)\theta_0(y)\theta_0(z) - \theta_1(x+y+z)\theta_2(x)\theta_2(y)\theta_2(z),\label{eqn:th8}
\end{equation}
which yields eq.~(\ref{eqn:th_ad1}) by putting $z=-y$ in eq.~(\ref{eqn:th8}). 
Other addition formulas are derived from eq.~(\ref{eqn:th_ad1}). 
Applying $x\rightarrow x+\frac{\tau}{3}$, $y\rightarrow y+\frac{\tau}{3}$ on eq.~(\ref{eqn:th_ad1}),
we have
\begin{equation}
 \theta_0(0)^2\theta_1(x+y)\theta_0(x-y) = \theta_0(x)\theta_1(x)\theta_1(y)^2-\theta_2(x)^2\theta_0(y)\theta_2(y).\label{eqn:a_th_ad2}
\end{equation}
Repeating the same procedure on eq.~(\ref{eqn:a_th_ad2}), we obtain
\begin{equation}
\theta_0(0)^2\theta_2(x+y)\theta_0(x-y) = \theta_0(x)\theta_2(x)\theta_0(y)^2-\theta_1(x)^2\theta_1(y)\theta_2(y).\label{eqn:a_th_ad3}
\end{equation}
Exchanging $x\leftrightarrow y$ in eq.~(\ref{eqn:th_ad1}), we have
\begin{equation}
 \theta_0(0)^2\theta_0(x+y)\theta_1(x-y) = \theta_0(x)\theta_1(x)\theta_0(y)^2-\theta_2(x)^2\theta_1(y)\theta_2(y).\label{eqn:a_th_ad4}
\end{equation}
Repeating $x\rightarrow x+\frac{\tau}{3}$, $y\rightarrow y+\frac{\tau}{3}$ on eq.~(\ref{eqn:a_th_ad4})
twice yield
\begin{eqnarray}
&& \theta_0(0)^2\theta_1(x+y)\theta_1(x-y)=\theta_0(x)\theta_2(x)\theta_2(y)^2-\theta_1(x)^2\theta_0(y)\theta_1(y),\label{eqn:a_th_ad5}\\
&& \theta_0(0)^2\theta_2(x+y)\theta_1(x-y)=\theta_1(x)\theta_2(x)\theta_1(y)^2-\theta_0(x)^2\theta_0(y)\theta_2(y),\label{eqn:a_th_ad6}
\end{eqnarray}
respectively. Applying $x\rightarrow x+\frac{\tau}{3}$ to eq.~(\ref{eqn:a_th_ad6}), we obtain
\begin{equation}
\theta_0(0)^2\theta_0(x+y)\theta_2(x-y)=\theta_0(x)\theta_2(x)\theta_1(y)^2-\theta_1(x)^2\theta_0(y)\theta_2(y).\label{eqn:a_th_ad7}
\end{equation}
Again, repeating $x\rightarrow x+\frac{\tau}{3}$, $y\rightarrow y+\frac{\tau}{3}$ on
eq.~(\ref{eqn:a_th_ad4}) twice, we have
\begin{eqnarray}
 &&\theta_0(0)^2\theta_1(x+y)\theta_2(x-y) = \theta_1(x)\theta_2(x)\theta_0(y)^2-\theta_0(x)^2\theta_1(y)\theta_2(y),\label{eqn:a_th_ad8}\\
 &&\theta_0(0)^2\theta_2(x+y)\theta_2(x-y) = \theta_0(x)\theta_1(x)\theta_2(y)^2-\theta_2(x)^2\theta_0(y)\theta_1(y),\label{eqn:a_th_ad9}
\end{eqnarray}
respectively. This completes the proof of Lemma \ref{lem:addition}.\qed


\begin{thebibliography}{99}
\bibitem{AD:Hesse_pencil} M.Artebani and I. Dolgachev. 
The Hesse pencil of plane cubic curves.
Preprint, arXiv:math/0611590v3 [math.AG], 2006.
\bibitem{BL:cubic} C. Birkenhake and H. Lange. 
Cubic theta relation. 
J. Reine Angew. Math. {\bf 407}(1990), 167--177.
\bibitem{Gathmann} A. Gathmann. 
Tropical algebraic geometry. 
Preprint, arXiv:math/0601322v1 [math.AG], 2006.
\bibitem{GRV} B. Grammaticos, A. Ramani and C.M. Viallet. 
Solvable chaos. 
Phys. Lett. {\bf A336}(2005), 152--158.
\bibitem{HHIK:AM_automata} G. Hatayama, K. Hikami, R. Inoue A. Kuniba, T. Takagi and T. Tokihiro. 
The $A_M^{(1)}$ automata related to cryctals of symmetric tensors.
J. Math. Phys. {\bf 42}(2001), 274--308.
\bibitem{HIK:crystal} K. Hikami, R. Inoue and Y. Komori. 
Crystallization	of the Bogoyavlensky lattice. 
J. Phys. Soc. Jpn. {\bf 68}(1999), 2234--2240.
\bibitem{HT:uTzitzeica} R. Hirota and D. Takahashi.
Ultradiscretization of the Tzitzeica equation.
Glasgow Math. J. {\bf 47}(2005), 77--85.
\bibitem{Ibukiyama} T. Ibukiyama. 
Modular forms of rational weights.
Introduction to modular	forms of half integral weights. Proceedings of the 8th summer
school on number theory, 2001, 1--22 (in Japanese).
\bibitem{Igusa} J. Igusa. 
Theta functions. 
Springer--Verlag, Berlin, 1972.
\bibitem{Inoue-Takenawa} R. Inoue and T. Takenawa.
Tropical spectral curves and integrable cellular automata.
Int. Math. Res. Notices  {\bf 2008}(2008), article ID rnn019.
\bibitem{IMNS:uSG} S. Isojima, M. Murata, A. Nobe and J. Satsuma.
An ultradiscretization of the sine--Gordon equation.
Phys. Lett. {\bf A331}(2004), 378--386.
\bibitem{IMS:tropical_book} I. Itenberg, G. Mikhalkin and E. Shustin.
Tropical algebraic geometry. 
Birkh\"auser, Basel, 2007.
\bibitem{Iwao-Tokihiro:upToda} S. Iwao and T. Tokihiro.
Ultradiscretization of the theta function solution of pd Toda.
J. Phys. A: Math. Theor. {\bf 40}(2007), 12987--13021.
\bibitem{Jacobi} C. G. J. Jacobi.
Theorie der Elliptischen Funktionen aus den Eigenscheften der Thetareihen abgeleitet.
Gesammelte Werke, Bd I, S. 497, G. Reimer, Berlin, 1881. 
\bibitem{JQ:Hessian} M. Joye and J.-J. Quisquater.
Hessian elliptic curves and side--channel attacks.
Cryptographic hardware and embedded systems--CHES 2001
(Lecture Notes in Computer Science, 2162). 
Eds \c{C}.K. Ko\c{c}, D. Naccache, and C. Paar. 
Springer--Verlag, Berlin, 2001, 402--410.
\bibitem{KNT:1d} K. Kajiwara, A. Nobe and T. Tsuda.
Ultradiscretization of solvable one--dimensional chaotic maps.
J. Phys. A: Math. Theor. {\bf 41}(2008), 395202(13pp).
\bibitem{KT:upToda} T. Kimijima and T. Tokihiro.
Initial--value problem of the discrete periodic Toda equation and its ultradiscretization.
Inverse Problems {\bf 18}(2002), 1705--1732.
\bibitem{Kirillov} A.N. Kirillov. 
Introduction to tropical combinatorics.  
Physics and combinatorics 2000. 
Eds A.N. Kirillov and N. Liskova. 
World Scientific, River Edge, NJ, 2001, 82--150.
\bibitem{KOSTY:KKR} A. Kuniba, M. Okado, R. Sakamoto, T. Takagi and Y. Yamada.
Crystal interpretation of Kerov--Kirillov--Reshetikhin bijection.
Nucl. Phys. {\bf B740}(2006), 299--327.
\bibitem{KS:utheta} A. Kuniba and R. Sakamoto.
Combinatorial Bethe ansatz and ultradiscrete Riemann theta function with rational characteristics.
Lett. Math. Phys. {\bf 80}(2007), 199--209.
\bibitem{KSY:tau_combinatorialBethe} A. Kuniba, R. Sakamoto and	Y. Yamada.
Tau functions in combinatorial Bethe ansatz.
Nucl. Phys. {\bf B786} (2007), 207--266.
\bibitem{KTT:Bethe} A. Kuniba, T. Takagi T and A. Takenouchi.
Bethe ansatz and inverse scattering transform in a periodic box--ball system.
Nucl. Phys. {\bf B747}(2006), 354--397.
\bibitem{MSTTT:uToda} J. Matsukidaira, J. Satsuma, D. Takahashi, T. Tokihiro and M. Torii.
Toda--type cellular automaton and its N--soliton solution.
Phys. Lett. {\bf A225}(1997), 287--295.
\bibitem{Mikhalkin:enumerative} G. Mikhalkin.
Enumerative tropical algebraic geometry in $\mathbb{R}^2$.
J. Amer. Math. Soc. {\bf 18}(2005), 313--377.
\bibitem{Mikhalkin:application} G. Mikhalkin.
Tropical geometry and its applications. 
Preprint, arXiv:math/0601041v2 [math.AG], 2006.
\bibitem{Mikhalkin-Zharkov} G. Mikhalkin and I. Zharkov. 
Tropical curves, their Jacobians and theta functions.
Preprint, arXiv:math/0612267v2 [math.AG], 2006.
\bibitem{MN:ELcorresponcende} J. Matsukidaira and K. Nishinari.
Euler--Lagrange correspondence of cellular automaton for traffic--flow models.
Phys. Rev. Lett. {\bf 90}(2003), 088701.
\bibitem{Mumford} D. Mumford.
Tata Lectures on Theta I.
Birkh\"auser, Boston, 1983.
\bibitem{N:uToda} H. Nagai.
A new expression of a soliton solution to the ultradiscrete Toda equation.
J. Phys. A: Math. Theor. {\bf 41}(2008), 235204 (12pp).
\bibitem{NMT:u-2dBuergers} K. Nishinari, J. Matsukidaira and D. Takahashi.
Two--dimensional Burgers cellular automaton.
J. Phys. Soc. Japan {\bf 70}(2001), 2267--2272.
\bibitem{NT:uBurgers} K. Nishinari and D. Takahashi.
Analytical properties of ultradiscrete Burgers equation and rule--184 cellular automaton.
J. Phys. A: Math. Gen: {\bf 31} (1998), 5439--5450.
\bibitem{Nobe:utheta} A. Nobe.
Ultradiscretization of elliptic functions and its applications to integrable systems.
J. Phys. A: Math. Gen. {\bf 39}(2006), L335--L342.
\bibitem{Nobe:uQRT} A. Nobe.
Ultradiscrete QRT maps and tropical elliptic curves.
J. Phys. A: Math. Theor. {\bf 41}(2008), 125205(12pp).
\bibitem{RST:1st_step} J. Richter-Gebert, B. Sturmfels and T. Theobald.
First steps in tropical	geometry.
Preprint, arXiv:math/0306366v2 [math.AG], 2003.
\bibitem{Smart:Hesse} N.P. Smart.
The Hessian form of an elliptic curve.
Cryptographic hardware and embedded systems--CHES 2001 (Lecture Notes in Computer Science, 2162).
Eds \c{C}.K. Ko\c{c}, D. Naccache, and C. Paar.
Springer--Verlag, Berlin, 2001, 118--125.
\bibitem{TH:Permanent} D. Takahashi and R. Hirota.
Ultradiscrete soliton solution of permanent type.
J. Phys. Soc. Jpn. {\bf 76}(2007), 104007.
\bibitem{TM:umKdV} D. Takahashi and J. Matsukdaira.
Box and ball system with a carrier and ultradiscrete modified KdV equation.
J. Phys. A: Math. Gen. {\bf 30}(1997), L733--L739.
\bibitem{TM:OVmodel} D. Takahashi and J. Matsukidaira.
On a discrete optimal velocity model and its continuous and ultradiscrete relatives.
Preprint, arXiv:0809.1265v1 [nlin.AO], 2008.
\bibitem{TS:SCA} D. Takahashi and J. Satsuma.
A soliton cellular automaton.
J. Phys. Soc. Japan {\bf 59}(1990), 3514--3519.
\bibitem{TSU:pattern_formation} D. Takahashi, A. Shida and M. Usami.
On the pattern formation mechanism of $(2+1)$D max--plus models.
J. Phys. A: Math. Gen. {\bf 34}(2001), 10715--10726.
\bibitem{TTMS:ultra} T. Tokihiro, D. Takahashi, J. Matsukidaira and J. Satsuma.
From soliton equations to integrable cellular automata through a limiting procedure.
Phys. Rev. Lett. {\bf 76}(1996), 3247--3250.
\bibitem{TTGOR:uP} D. Takahashi, T. Tokihiro, B. Grammaticos, Y. Ohta and A. Ramani.
Constructing solutions to the ultradiscrete Painlev\'e equations.
J. Phys. A: Math. Gen. {\bf 30}(1997), 7953--7966.
\bibitem{TM:bb_and_Riemann} T. Tokihiro and J. Mada.
Fundamental cycle of a periodic box--ball system: a number theoretical aspect.
Glasgow Math. J. {\bf 47}(2005), 199--204.
\bibitem{TTM:undKP} T. Tokihiro T, D. Takahashi and J. Matsukidaira.
Box and ball system as a realization of ultradiscrete nonautonomous KP equation.
J. Phys. A: Math. Gen. {\bf 33}(2000), 607--619.
\bibitem{Tsuda:tropicalWeyl} T. Tsuda.
Tropical Weyl group action via point configurations and $\tau$--functions of the $q$--Painlev\'e equations.
Lett. Math. Phys. {\bf 77} (2006), 21--30.
\bibitem{TT:tropicalWeyl} T. Tsuda and T. Takenawa.
Tropical representation of Weyl groups	associated with certain rational varieties.
Adv. in Math. (in press).
\bibitem{TH:uKdV} S. Tsujimoto and R. Hirota.
Ultradiscrete KdV equation.
J.Phys. Soc. Jpn. {\bf 67}(1998), 1809--1810.
\bibitem{Vigeland} M.D. Vigeland.
The group law on a tropical elliptic curve.
Preprint, arXiv:math/0411485v1 [math.AG], 2004.
\bibitem{Y:Amoebae} Y. Yamada.
An illustrated guide to the amoebae of type E --collection and observation--.
Talk delivered at colloquium of Institute of Physics, 
Graduate School of Liberal Arts and Sciences, the University of Tokyo, 2007.
\bibitem{YYT:pBBS} D. Yoshihara, F. Yura and T. Tokihiro.
Fundamental cycle of a periodic box--ball system.
J. Phys. A: Math. Gen. {\bf 36}(2003), 99--121.
\end{thebibliography}
\end{document}